\documentclass[twocolumn]{article}
\usepackage[utf8]{inputenc}
\usepackage[backend=biber,style=nature,sorting=none,isbn=false,clearlang=true,
  url=false,doi=false,eprint=false,hyperref=true,punctfont=false]{biblatex}
\usepackage{preprint-twocol}
\usepackage{balance}
\AtEveryBibitem{\ifnumequal{\value{refsection}}{0}{\iffieldequalstr{labelnumber}{54}{\balance}{}}{}}
\usepackage{hyperref}
\hypersetup{pdftitle={Two channels of magnetic renormalization in the terahertz lattice response of FePS3},
  pdfauthor={Eduardo D. Stefanato, Nicolas M. Kawahala, Ana Carolina F. de Brito, Jan Plutnar, Zdenek Sofer, Bernardo R. A. Neves, Ingrid D. Barcelos, Felix G. G. Hernandez}}

\AtEveryBibitem{\clearfield{month}}
\AtEveryBibitem{\clearlist{language}}
\AtEveryBibitem{\clearfield{note}}
\AtEveryBibitem{\clearfield{issue}}
\DeclareSourcemap{
  \maps[datatype=bibtex]{
    \map[overwrite=true]{
      \step[fieldsource=journal, match=\regexp{\A\s*2D\s+Materials\s*\Z}, replace={2D Mater.}]
      \step[fieldsource=journal, match=\regexp{\A\s*Advanced\s+Materials\s*\Z}, replace={Adv. Mater.}]
      \step[fieldsource=journal, match=\regexp{\A\s*ACS\s+Applied\s+Electronic\s+Materials\s*\Z}, replace={ACS Appl. Electron. Mater.}]
      \step[fieldsource=journal, match=\regexp{\A\s*Advances\s+in\s+Physics\s*\Z}, replace={Adv. Phys.}]
      \step[fieldsource=journal, match=\regexp{\A\s*Brazilian\s+Journal\s+of\s+Physics\s*\Z}, replace={Braz. J. Phys.}]
      \step[fieldsource=journal, match=\regexp{\A\s*Helvetica\s+Physica\s+Acta\s*\Z}, replace={Helv. Phys. Acta}]
      \step[fieldsource=journal, match=\regexp{\A\s*Journal\s+of\s+Applied\s+Physics\s*\Z}, replace={J. Appl. Phys.}]
      \step[fieldsource=journal, match=\regexp{\A\s*Journal\s+of\s+Infrared,\s+Millimeter,\s+and\s+Terahertz\s+Waves\s*\Z}, replace={J. Infrared Millim. Terahertz Waves}]
      \step[fieldsource=journal, match=\regexp{\A\s*Journal\s+of\s+Magnetism\s+and\s+Magnetic\s+Materials\s*\Z}, replace={J. Magn. Magn. Mater.}]
      \step[fieldsource=journal, match=\regexp{\A\s*Journal\s+of\s+the\s+Optical\s+Society\s+of\s+America\s+B\s*\Z}, replace={J. Opt. Soc. Am. B}]
      \step[fieldsource=journal, match=\regexp{\A\s*Journal\s+of\s+the\s+Physical\s+Society\s+of\s+Japan\s*\Z}, replace={J. Phys. Soc. Jpn.}]
      \step[fieldsource=journal, match=\regexp{\A\s*Journal\s+of\s+the\s+American\s+Statistical\s+Association\s*\Z}, replace={J. Am. Stat. Assoc.}]
      \step[fieldsource=journal, match=\regexp{\A\s*Laser\s+\S+\s+Photonics\s+Reviews\s*\Z}, replace={Laser Photonics Rev.}]
      \step[fieldsource=journal, match=\regexp{\A\s*Materials\s+Research\s+Bulletin\s*\Z}, replace={Mater. Res. Bull.}]
      \step[fieldsource=journal, match=\regexp{\A\s*Nano\s+Letters\s*\Z}, replace={Nano Lett.}]
      \step[fieldsource=journal, match=\regexp{\A\s*Nature\s+Communications\s*\Z}, replace={Nat. Commun.}]
      \step[fieldsource=journal, match=\regexp{\A\s*Nature\s+Mater(ials)?\s*\Z}, replace={Nat. Mater.}]
      \step[fieldsource=journal, match=\regexp{\A\s*Nature\s+Nanotechnology\s*\Z}, replace={Nat. Nanotechnol.}]
      \step[fieldsource=journal, match=\regexp{\A\s*Nature\s+Physics\s*\Z}, replace={Nat. Phys.}]
      \step[fieldsource=journal, match=\regexp{\A\s*Nature\s+Reviews\s+Physics\s*\Z}, replace={Nat. Rev. Phys.}]
      \step[fieldsource=journal, match=\regexp{\A\s*npj\s+2D\s+Materials\s+and\s+Applications\s*\Z}, replace={npj 2D Mater. Appl.}]
      \step[fieldsource=journal, match=\regexp{\A\s*npj\s+Quantum\s+Materials\s*\Z}, replace={npj Quantum Mater.}]
      \step[fieldsource=journal, match=\regexp{\A\s*Optical\s+Materials\s*\Z}, replace={Opt. Mater.}]
      \step[fieldsource=journal, match=\regexp{\A\s*Physical\s+Review\s*\Z}, replace={Phys. Rev.}]
      \step[fieldsource=journal, match=\regexp{\A\s*Physical\s+Review\s+B\s*\Z}, replace={Phys. Rev. B}]
      \step[fieldsource=journal, match=\regexp{\A\s*Physical\s+Review\s+Letters\s*\Z}, replace={Phys. Rev. Lett.}]
      \step[fieldsource=journal, match=\regexp{\A\s*Physical\s+Review\s+Materials\s*\Z}, replace={Phys. Rev. Mater.}]
      \step[fieldsource=journal, match=\regexp{\A\s*Physical\s+Review\s+Research\s*\Z}, replace={Phys. Rev. Res.}]
      \step[fieldsource=journal, match=\regexp{\A\s*Physical\s+Review\s+X\s*\Z}, replace={Phys. Rev. X}]
      \step[fieldsource=journal, match=\regexp{\A\s*Physics\s+Letters\s+A\s*\Z}, replace={Phys. Lett. A}]
      \step[fieldsource=journal, match=\regexp{\A\s*Physics\s+Reports\s*\Z}, replace={Phys. Rep.}]
      \step[fieldsource=journal, match=\regexp{\A\s*Proceedings\s+of\s+the\s+National\s+Academy\s+of\s+Sciences\s*\Z}, replace={Proc. Natl Acad. Sci. USA}]
      \step[fieldsource=journal, match=\regexp{\A\s*Proceedings\s+of\s+the\s+Royal\s+Society\s+of\s+London\s+A\s*\Z}, replace={Proc. R. Soc. Lond. A}]
      \step[fieldsource=journal, match=\regexp{\A\s*Progress\s+of\s+Theoretical\s+Physics\s*\Z}, replace={Prog. Theor. Phys.}]
      \step[fieldsource=journal, match=\regexp{\A\s*Reviews\s+of\s+Modern\s+Physics\s*\Z}, replace={Rev. Mod. Phys.}]
      \step[fieldsource=journal, match=\regexp{\A\s*Science\s+Advances\s*\Z}, replace={Sci. Adv.}]
      \step[fieldsource=journal, match=\regexp{\A\s*Solid\s+State\s+Ionics\s*\Z}, replace={Solid State Ion.}]
      \step[fieldsource=journal, match=\regexp{\A\s*The\s+Annals\s+of\s+Mathematical\s+Statistics\s*\Z}, replace={Ann. Math. Stat.}]
      \step[fieldsource=journal, match=\regexp{\A\s*The\s+Annals\s+of\s+Statistics\s*\Z}, replace={Ann. Stat.}]
    }
    \map{
      \step[fieldsource=title, match=\regexp{\$_\{?(\d+)\}?\$}, replace=\regexp{\\textsubscript\{$1\}}]
    }
  }
}

\defbibheading{bibliography}[References]{\section*{#1}}
\newbibmacro{pp:doilink}[1]{%
  \iffieldundef{doi}{#1}{{\hypersetup{urlcolor=black}\href{https://doi.org/\thefield{doi}}{#1}}}}
\DeclareFieldFormat[article]{title}{\usebibmacro{pp:doilink}{#1}}
\DeclareFieldFormat[misc]{title}{#1}
\DeclareBibliographyDriver{misc}{%
  \usebibmacro{bibindex}%
  \usebibmacro{begentry}%
  \usebibmacro{author/editor+others/translator+others}%
  \setunit{\labelnamepunct}\newblock
  \usebibmacro{title}%
  \newunit\newblock
  \ifboolexpr{test {\iffieldequalstr{eprinttype}{arXiv}} or test {\iffieldequalstr{eprinttype}{arxiv}}}
    {\printtext{Preprint at \url{https://arxiv.org/abs/\thefield{eprint}}}}
    {\printfield{howpublished}%
     \setunit{\addspace}\printlist{publisher}%
     \setunit{\addspace}\printfield{doi}}%
  \setunit{\addspace}%
  \printtext[parens]{\usebibmacro{date}}%
  \usebibmacro{finentry}}
\DeclareCiteCommand{\citenum}
  {\usebibmacro{cite:init}}
  {\usebibmacro{citeindex}\usebibmacro{cite:comp}}
  {}
  {\usebibmacro{cite:dump}}

\renewcommand\pptitle{Two channels of magnetic renormalization in the terahertz lattice response of FePS$_{3}$}
\renewcommand\ppauthors{\mbox{Eduardo D. Stefanato$^{1}$}, \mbox{Nicolas M. Kawahala$^{1}$}, \mbox{Ana Carolina F. de Brito$^{2}$}, \mbox{Jan Plutnar$^{3}$}, \mbox{Zdenek Sofer$^{3}$}, \mbox{Bernardo R. A. Neves$^{4}$}, \mbox{Ingrid D. Barcelos$^{2}$} and \mbox{Felix G. G. Hernandez$^{1,\dagger}$}}
\renewcommand\ppaffiliations{$^1$Instituto de Física, Universidade de São Paulo, São Paulo, SP, 05508-090, Brazil.\\
$^2$Brazilian Synchrotron Light Laboratory, Brazilian Center for Research in Energy and Materials, Campinas, SP, 13083-100, Brazil.\\
$^3$Dept. of Inorganic Chemistry, University of Chemistry and Technology Prague, Technicka 5, 166 28 Prague 6, Czech Republic.\\
$^4$Physics Department, Instituto de Ciências Exatas, Universidade Federal de Minas Gerais, Belo Horizonte, MG, 31270-901, Brazil.\\
$^\dagger$Corresponding author. Email: \href{mailto:felixggh@if.usp.br}{felixggh@if.usp.br}}
\renewcommand\ppabstract{Interlayer vibrations probe how magnetic van der Waals layers are stacked and coupled. In the zigzag antiferromagnet FePS$_3$, ultrafast experiments show that the interlayer shear couples strongly to the magnetic order, but not how magnetism renormalizes the shear stiffness and damping. Here, using terahertz time-domain spectroscopy, we measure the anisotropic conductivity of FePS$_3$ from 1.6 to 300~K and in fields up to 7~T normal to the layers. We observe absorption bands at 0.6--0.8 and 1.3--2.4~THz that persist into the paramagnetic phase. The lower one is consistent with an interlayer shear resonance; the magnetic order stiffens one of its in-plane components and softens the other, reversing their splitting at the N\'eel temperature. We identify two channels of magnetic renormalization: frequency shifts are confined to the ordered phase, whereas the linewidths of this resonance and most oscillator strengths follow short-range spin correlations persisting well above the transition. The field acts mainly on the higher resonance in one orientation: near 3.2~T, its softening accelerates and its linewidth has a cusp-like minimum, consistent with critical slowing down of a field-induced order parameter. These results give an equilibrium view of spin--shear coupling, which is anisotropic and persists above the N\'eel temperature through short-range correlations.}
\renewcommand\pprunningauthors{Stefanato et al.}

\begin{document}
\makepptitle

\section*{Introduction}\label{sec:intro}
Layered van der Waals (vdW) magnets retain long-range magnetic order down to a few or even a single atomic layer \cite{Huang2017,Gong2017,Lee2016, Burch2018, Gibertini2019, GongZhang2019}. Their layers are held together by weak interlayer forces, so that the stacking sequence, twinning and interlayer vibrations become part of the physics of these materials~\cite{Sivadas2018, Song2019}, and low-frequency interlayer modes are sensitive probes of how the layers are arranged and coupled \cite{Tan2012,ZhangX2013, Liang2017}. In vdW antiferromagnets, these lattice degrees of freedom coexist with strongly anisotropic magnetic order, and the coupling between the two can be read from the energies, lifetimes and optical activity of the low-energy excitations.

FePS$_3$, a member of the transition-metal thiophosphate family MPS$_3$~\cite{Brec1986,Susner2017}, is a prototypical example. Its Fe$^{2+}$ ions form honeycomb layers~\cite{Ouvrard1985} that order below $T_\mathrm{N}\approx118$~K~\cite{Kurosawa1983} in a zigzag antiferromagnetic structure, with the moments nearly normal to the
layers~\cite{Lancon2016,Lee2016}. A strong single-ion anisotropy and an interlayer exchange two orders of magnitude weaker than the intralayer one make it a quasi-two-dimensional Ising antiferromagnet~\cite{Jernberg1984,Joy1992,Rule2007,Lancon2016,Liu2021,ZhangQ2021}, which remains ordered down to the monolayer \cite{Lee2016,Wang2016}. The magnetic ordering is accompanied by a lattice distortion \cite{Murayama2016}, and bulk crystals contain rotational twins whose boundaries lie in the vdW gaps \cite{Murayama2016,Lancon2016}. The zigzag order also gives rise to a giant optical linear dichroism, which persists down to the monolayer and probes zigzag domains oriented at 60$^\circ$ to one another~\cite{ZhangQ2021LD}. Under pressure, adjacent layers slide into a different stacking~\cite{Haines2018}, and the magnetic coupling between them changes from antiferromagnetic to ferromagnetic~\cite{Coak2021}. The spin--orbit-coupled Fe$^{2+}$ ion gives rise to magnetic excitations of mixed spin and orbital character~\cite{Dhakal2024}, the lowest of which is the zone-centre magnon near 3.7~THz (122~cm$^{-1}$)~\cite{Sekine1990,McCreary2020,Vaclavkova2021}.

Spin--lattice coupling in FePS$_3$ is strong and markedly mode selective~\cite{Ghosh2021}. Below $T_\mathrm{N}$, zone-boundary phonons folded to the zone centre by the zigzag order appear between 2.6 and 3.3~THz
(88--108~cm$^{-1}$)~\cite{Scagliotti1987,Sekine1990,Lee2016,Wang2016,McCreary2020}, and in fields above $\approx$10~T they hybridize with the magnon into magnon polarons \cite{Vaclavkova2021,Liu2021,ZhangQ2021}. Coherent-phonon experiments show that phonon amplitudes and phases follow the magnetic order, in some modes without any frequency renormalization \cite{Ergecen2023,Mertens2023}. The interlayer shear of FePS$_3$ is itself coupled to the zigzag order: its recovery after photoexcitation slows down critically near $T_\mathrm{N}$ \cite{ZhangXX2021,Zhou2022}, and spin-mediated shear oscillations have been reported \cite{Zong2023}. Intense THz fields have further been used to induce a metastable magnetization near criticality \cite{Ilyas2024} and to drive linear and nonlinear magnon--phonon dynamics \cite{Luo2025,Li2026}.

These excitations lie almost entirely above $\approx$2.6~THz, and the linear electrodynamic response below this frequency has been accessed only partially. Far-infrared magneto-transmission spectra are commonly divided by their average over field  \cite{Liu2021,Wyzula2022,LeMardel2024}, a representation designed to isolate field-dependent excitations that suppresses field-independent absorption, and aperture-referenced magneto-transmission starts at 1.8~THz (60~cm$^{-1}$) \cite{Vaclavkova2021}. Absolute infrared transmission and absorption spectra have been reported down to 1.2--1.5~THz (40--50~cm$^{-1}$), in samples 10--26~$\mu$m
thick \cite{JouanneJulien1988,Panda2026}. Raman scattering reaches
$\approx$0.3~THz (10~cm$^{-1}$) \cite{Sekine1990}, but in the centrosymmetric $C2/m$ structure the Raman- and infrared-active zone-centre phonons are mutually exclusive \cite{Wang2016,McCreary2020,Panda2026}, and nonlinear THz experiments detect difference-frequency signals rather than linear
absorption \cite{Li2026}. To our knowledge, neither the linear absorption of FePS$_3$ below $\approx$1.2~THz nor its complex conductivity obtained from the amplitude and phase of the transmitted field, across $T_\mathrm{N}$ and in magnetic field, has been reported. This low-frequency range is where interlayer shear and breathing vibrations of layered crystals are expected \cite{Tan2012,ZhangX2013}, and where the coupling between the interlayer lattice and the magnetic order identified in FePS$_3$ \cite{Zhou2022,Zong2023} may leave its signature in the optical response.

Here we report the low-energy electrodynamics of bulk FePS$_3$ measured by magneto THz time-domain spectroscopy, which yields the complex optical conductivity from the amplitude and phase of the transmitted field \cite{Grischkowsky1990, Lloyd-Hughes2012-es, marulanda2026colossal, Kawahala2026}. The measurements cover 0.15--2.65~THz, temperatures from 1.6 to 300~K, two orthogonal in-plane orientations of a 298$\pm$1${\mu}$m thick-crystal, and magnetic fields up to 7~T applied normal to the layers. We resolve a pair of resonances near 0.72 and 0.76~THz (24 and 25~cm$^{-1}$) at low temperature and two higher modes, near 2.0--2.1 and 2.35~THz (67--70 and 78~cm$^{-1}$). The lower pair is consistent with the two in-plane components of an interlayer shear resonance made infrared active by stacking disorder, whereas the higher modes behave as distinct lattice vibrations. Across $T_\mathrm{N}$, the magnetic order renormalizes the frequencies, linewidths and oscillator strengths in a mode- and polarization-selective way, which we describe with the squared order parameter and the nearest-neighbour spin correlation of the two-dimensional Ising model on the honeycomb lattice. In magnetic field, the higher mode in one of the two orientations changes its frequency, linewidth and oscillator strength at a common field, $B^*\approx3.2$~T, an order of magnitude below the field-induced transitions known for FePS$_3$ \cite{Liu2021,Wildes2020} and in a range where its static magnetization shows no transition \cite{Wildes2020}. The magnetic order acts on the THz lattice response through two channels: the magnetic frequency shifts follow the long-range order and are confined to the ordered phase, whereas the linewidths of the lower pair and most oscillator strengths follow short-range spin correlations that persist well above $T_\mathrm{N}$. Linear THz spectroscopy thus complements, at equilibrium, the time-resolved experiments that revealed the coupling between interlayer shear and magnetic
order in this material~\cite{Zhou2022,Zong2023,Zhou2026}.

\section*{Results and discussion}\label{sec:results}

\begin{figure}[t]\begin{ppfloat}
\centering
\includegraphics[width=80.1mm]{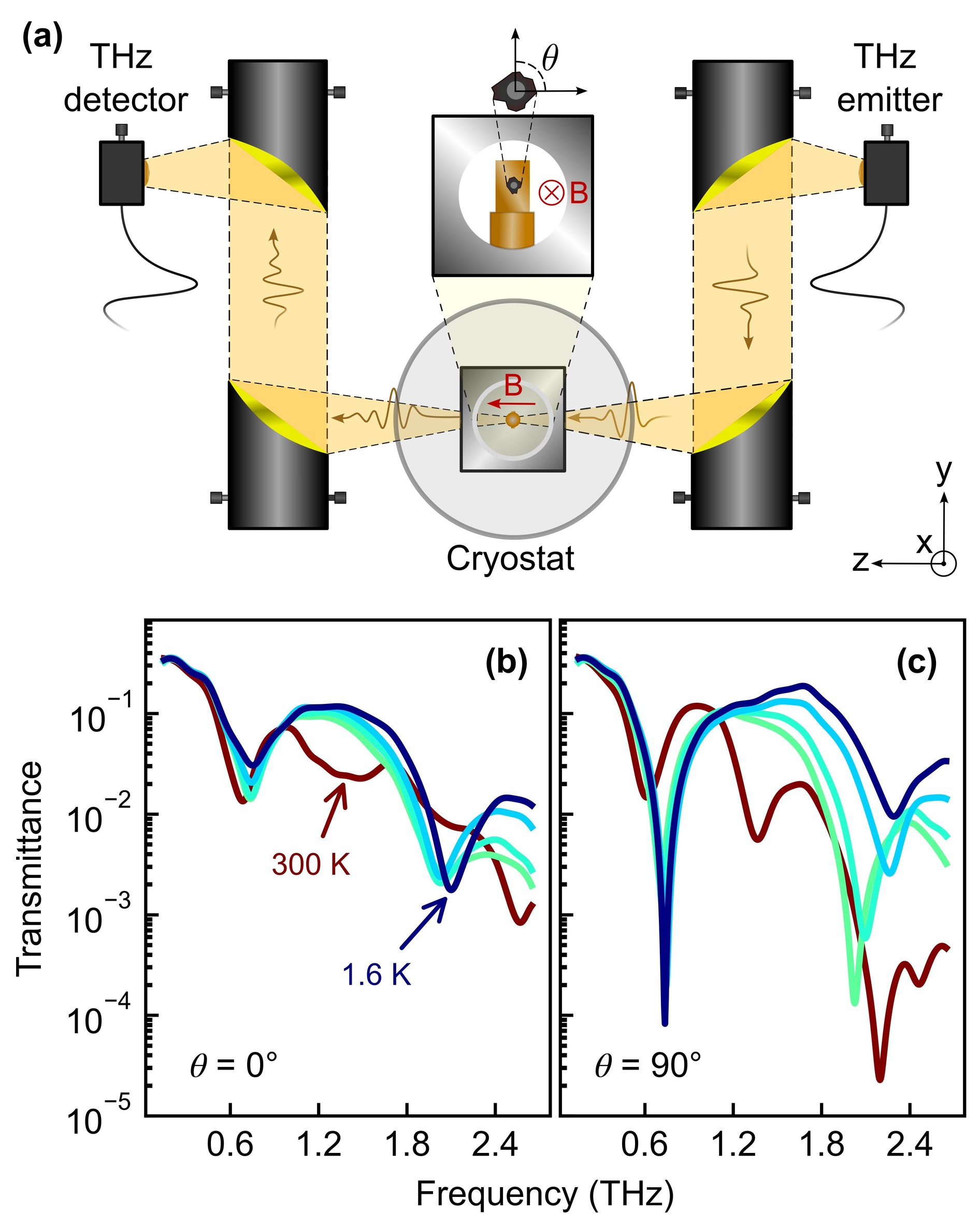}
\caption{\textbf{Experimental geometry and temperature-dependent THz transmittance of FePS$_3$.} \textbf{a} Schematic of the transmission THz-TDS set-up. Off-axis parabolic mirrors focus the THz pulse from the emitter onto the crystal, mounted in a magneto-optical cryostat, and collect the transmitted pulse onto the detector. The magnetic field is applied along the propagation direction, normal to the layers ($\mathbf{B}\parallel c^*\parallel\mathbf{k}_\mathrm{THz}$; Faraday geometry). The angle $\theta$ denotes a rotation of the crystal about $c^*$, which changes its in-plane orientation relative to the fixed THz polarization; $\theta$ is an experimental label and is not referred to a crystallographic axis. \textbf{b},~\textbf{c} Intensity transmittance at zero field for $\theta=0^\circ$ and $\theta=90^\circ$, on a logarithmic scale and referenced to a sample-free measurement, at 1.6, 100, 120, 140 and 300~K.}\label{fig:transmission}
\end{ppfloat}\end{figure}

Figure~\ref{fig:transmission}a shows the geometry of the experiment, the THz pulse propagates normal to the layers, parallel to the applied magnetic field. The angles $\theta=0^\circ$ and $90^\circ$ denote two sample orientations related by a $90^\circ$ rotation about $c^*$ relative to the fixed THz polarization. The in-plane crystallographic axes were not independently indexed in this mounting, so the angle between the THz eletric field and the zigzag chains is unknown. Figures~\ref{fig:transmission}b,c show the zero-field transmittance for these two orientations. Both contain two absorption bands below 2.5~THz: a lower band near 0.6--0.8~THz and a broader band between $\approx$1.3 and 2.4~THz. At low temperature, the transmittance at the band centres falls to $10^{-2}$--$10^{-4}$. The positions, widths and thermal evolution of the bands change on rotating the crystal by $90^\circ$. The higher band shows the largest difference: at $\theta=90^\circ$, it shifts from $\approx$2.35~THz at low temperature to $\approx$1.35~THz at 300~K, whereas at $\theta=0^\circ$ it remains near 2.0~THz and weakens on warming.
 
\begin{figure*}[t]\begin{ppfloat}
\centering
\includegraphics[width=154.3mm]{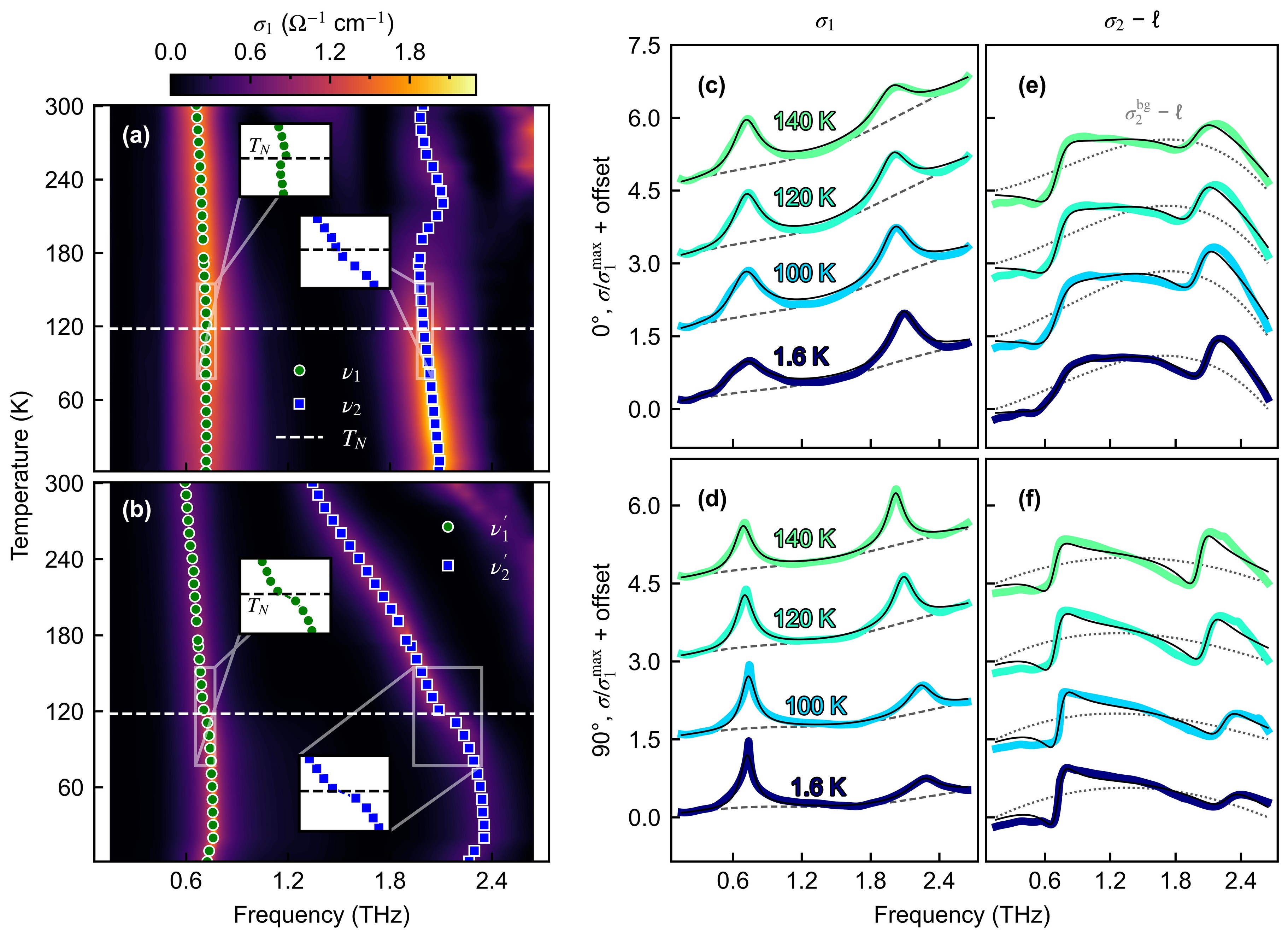}
\caption{\textbf{Temperature evolution of the anisotropic THz conductivity of FePS$_3$.} \textbf{a},~\textbf{b} Colour maps of the real part of the optical conductivity after subtraction of a smooth baseline, $|\sigma_1-\sigma_b|$, as a function of frequency and temperature for $\theta=0^\circ$ and $\theta=90^\circ$ (Methods). Circles and squares mark the resonance frequencies of the lower (mode~1; $\nu_1$, $\nu_1'$) and higher (mode~2; $\nu_2$, $\nu_2'$) modes obtained from the fits; primes denote $\theta=90^\circ$. Dashed lines mark $T_\mathrm{N}$, and the insets enlarge the boxed regions around $T_\mathrm{N}$. \textbf{c},~\textbf{d} Real part $\sigma_1$ and \textbf{e},~\textbf{f} imaginary part $\sigma_2-\ell$ of the conductivity at 1.6, 100, 120 and 140~K, for $\theta=0^\circ$ (\textbf{c}, \textbf{e}) and $\theta=90^\circ$ (\textbf{d}, \textbf{f}). Each spectrum is divided by $\sigma_1^\mathrm{max}$, the maximum of the sum of the two fitted oscillators in $\sigma_1$, and offset vertically for clarity. Coloured lines are the data. In \textbf{c} and \textbf{d}, dashed lines are the baseline $\sigma_b$, and black lines are $\sigma_b$ plus the two damped harmonic oscillators obtained from simultaneous fits of $\sigma_1$ and $\sigma_2$. In \textbf{e} and \textbf{f}, dotted lines are a smooth non-resonant background of $\sigma_2$, $\sigma_2^\mathrm{bg}$, and black lines add to it the imaginary parts of the same oscillators; $\ell$, the straight line that joins $\sigma_2^\mathrm{bg}$ at the ends of the measured range (0.15 and 2.65~THz), is subtracted from all curves for display only.}\label{fig:conductivity}
\end{ppfloat}\end{figure*}

Two features of these spectra already constrain the nature of the bands. First, both persist in the paramagnetic phase, the lower one up to 300~K in both orientations. They are therefore not excitations of the ordered phase, such as the magnon and the zone-folded phonons of the zigzag phase, which appear only below $T_\mathrm{N}$ and lie above 2.6~THz~\cite{Lee2016,McCreary2020}. Second, the anisotropy persists above $T_\mathrm{N}$. Each layer of FePS$_3$ has threefold rotational symmetry, and the in-plane anisotropy of the paramagnetic crystal derives from the monoclinic stacking, in which successive layers are
shifted along $a$~\cite{Lee2016}. In Raman and infrared spectra, this anisotropy is weak, and the phonons of FePS$_3$ become markedly anisotropic only below $T_\mathrm{N}$~\cite{Lee2016,Panda2026}. The lower THz band, in contrast, remains clearly anisotropic at 300~K (Fig.~\ref{fig:conductivity}), as expected for a vibration that involves the relative motion of the layers and therefore probes
the stacking directly; its frequency, moreover, falls in the range expected for such interlayer vibrations~\cite{Tan2012,ZhangX2013}. We return to these points in the last subsection, where we discuss the origin of the resonances.
 
\subsection*{Low-energy resonances in the optical conductivity}\label{sec:conductivity} 
 
In the optical conductivity, each absorption band appears as one resonance in each orientation (Fig.~\ref{fig:conductivity}a,b). Each resonance is described by a damped harmonic oscillator with frequency $\nu$, linewidth $\gamma$ and oscillator strength $A$, and the same oscillators reproduce $\sigma_1$, in which the resonances appear as peaks, and $\sigma_2$, in which they appear as dispersive features (Fig.~\ref{fig:conductivity}c--f; Methods). At 20~K,
$\nu_1\approx0.72$, $\nu_1'\approx0.76$, $\nu_2\approx2.1$ and
$\nu_2'\approx2.4$~THz (24, 25, 70 and 79~cm$^{-1}$). Mode~1 is strongly damped at $\theta=0^\circ$, where $\gamma_1\approx0.38$~THz is about half its frequency, and less than half as wide at $\theta=90^\circ$ ($\gamma_1'\approx0.16$~THz).
 
\begin{figure*}[t]\begin{ppfloat}
\centering
\includegraphics[width=155.4mm]{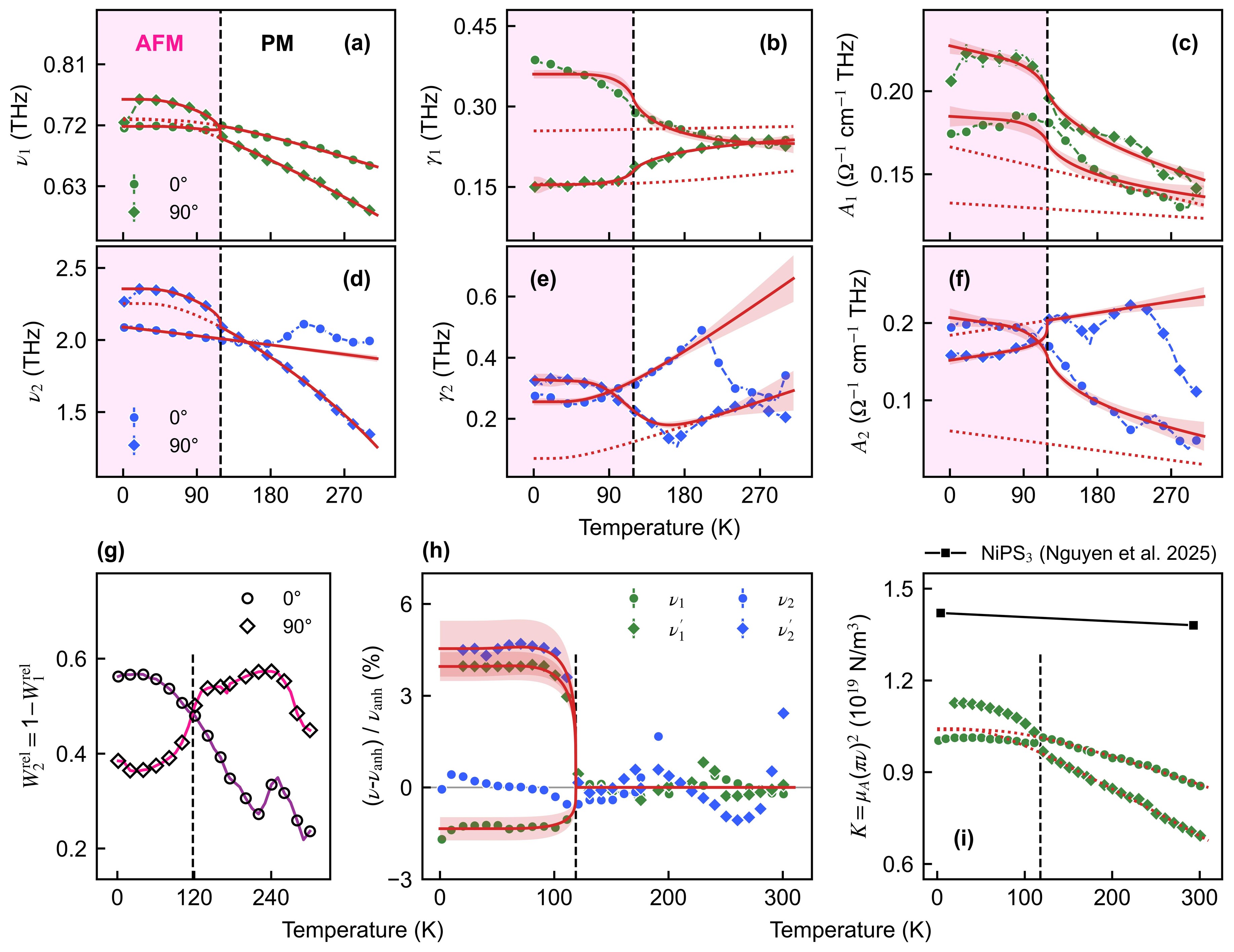}
\caption{\textbf{Temperature dependence of the resonance parameters and spin--lattice renormalization.} \textbf{a},~\textbf{d} Resonance frequencies,~\textbf{b},~\textbf{e} linewidths (FWHM) and \textbf{c},~\textbf{f} oscillator strengths of the lower ($\nu_1$, $\nu_1'$; green) and higher ($\nu_2$, $\nu_2'$; blue) modes for $\theta=0^\circ$ (circles) and $\theta=90^\circ$ (diamonds), obtained from complex fits of the THz conductivity with one damped harmonic oscillator per mode. Symbols are shown for every second temperature, and thin dashed lines join all measured points. Error bars are the uncertainties of the fits; at 161 and 176~K ($\theta=90^\circ$), those of mode~2 include a systematic contribution (Methods). The shaded region marks the antiferromagnetic (AFM) phase, and dashed vertical lines mark $T_\mathrm{N}$. Solid red lines are fits with an anharmonic background plus a spin--lattice term proportional to $\Phi(T)$, the squared order parameter or the nearest-neighbour spin correlation of the two-dimensional Ising model on the honeycomb lattice (for $\gamma_2'$, a phenomenological step); dotted red lines are the same fits without the spin--lattice term, where it is retained, and shaded red areas are 95\% confidence bands. Model forms and fitted ranges are given in Methods and the selection criteria in Supplementary Note~1; outside the fitted ranges, the curves are extrapolations. \textbf{g} Relative spectral weight of the higher mode, $W_2^\mathrm{rel}=W_2/(W_1+W_2)=1-W_1^\mathrm{rel}$, where $W_i$ is the integral of the fitted oscillator of mode $i$ over $\nu_i\pm0.30$~THz; lines join all temperatures, and symbols are shown for every second temperature. \textbf{h} Relative magnetic contribution to the frequencies, $(\nu-\nu_\mathrm{anh})/\nu_\mathrm{anh}$, where $\nu_\mathrm{anh}$ is the anharmonic background of the fits in \textbf{a} and \textbf{d}; solid lines, $\lambda\Phi/\nu_\mathrm{anh}$. For $\nu_2$, which has no spin--lattice term, the points are the relative residuals of the fit. \textbf{i} Lower bound to the interlayer shear force constant per unit area, $K=\mu_A(\pi\nu)^2$, obtained from $\nu_1$ and $\nu_1'$ for a linear chain of rigid layers with mass per unit area $\mu_A=1.985\times10^{-6}$~kg\,m$^{-2}$; dotted lines, $K$ calculated with $\nu_\mathrm{anh}$; black squares, the shear force constant of NiPS$_3$\cite{Nguyen2025}.}\label{fig:parameters}
\end{ppfloat}\end{figure*}

The magnetic order leaves clear marks on these resonances. Around $T_\mathrm{N}$ (insets of Fig.~\ref{fig:conductivity}a,b), $\nu_1'$ and $\nu_2'$ fall by $\approx$23 and $\approx$100~GHz on warming from 111 to 121~K, two to three times more than in the adjacent 10-K intervals, whereas $\nu_1$ rises by $\approx$6~GHz, against its thermal softening, and $\nu_2$ evolves smoothly. Because $\nu_1$ and $\nu_1'$ move in opposite directions~\cite{Kant2012}, the splitting of mode~1, $\nu_1'-\nu_1$, changes sign at $T_\mathrm{N}$: it is positive in the
ordered phase, 39~GHz at 20~K, and negative above, down to $-66$~GHz at 300~K. The line shapes change across $T_\mathrm{N}$ as well
(Fig.~\ref{fig:conductivity}c--f). Between 100 and 140~K, mode~1 broadens and mode~2 narrows at $\theta=90^\circ$, whereas at $\theta=0^\circ$ mode~1 narrows and mode~2 broadens and weakens; above $T_\mathrm{N}$, mode~2 at $\theta=0^\circ$ keeps broadening and weakening.
 
These anomalies show that the probed modes are coupled to the magnetic order, and that the coupling depends on both the mode and the in-plane direction. For mode~1, the frequency shifts of $\nu_1$ and $\nu_1'$ exhibit opposite signs in the two measured orientations. This is consistent with an anisotropic renormalization of the in-plane shear stiffness, determined by the zigzag order~\cite{Lancon2016,Panda2026}, and with the in-plane components of a single vibration. Which shift corresponds to shear along the chains remains undetermined. For mode~2, the anomaly at $\theta=90^\circ$ and its absence at $\theta=0^\circ$ suggest instead that $\nu_2$ and $\nu_2'$ need not belong to the same vibration. Figure~\ref{fig:parameters} quantifies these couplings, and we return to the identity of the modes in the last subsection.
 
\subsection*{Temperature dependence and spin--lattice renormalization}\label{sec:temperature}
 
To separate the effect of the magnetic order from the anharmonic renormalization of the lattice, we add to the anharmonic background of each parameter in Fig.~\ref{fig:parameters}a--f a term proportional to a spin correlation function, $\Phi(T)$, normalized to unity at $T=0$ (Methods), the standard description of the renormalization of phonons by magnetic order \cite{Baltensperger1968,Granado1999,Sushkov2005}. We denote the amplitude of this term by $\lambda$, $\eta$ and $\zeta$ for the frequencies, linewidths and oscillator strengths. Two forms of $\Phi$ distinguish a response to long-range order from one to short-range correlations: the squared order parameter, $\Phi_\mathrm{LRO}$, which vanishes above $T_\mathrm{N}$, and the nearest-neighbour spin correlation, $\Phi_\mathrm{SRO}$, which persists above it. Since FePS$_3$ is a quasi-two-dimensional Ising antiferromagnet with the Fe$^{2+}$ ions on a honeycomb lattice, both functions are taken from the corresponding Ising model~\cite{Rule2007}. They do not describe the zigzag structure, which involves competing exchange interactions up to third neighbours \cite{Lancon2016}, but they capture long- and short-range order with a single temperature scale, $T_\mathrm{N}$. The spin--lattice term is retained only where the data require it, and for each parameter we keep the form of $\Phi$ that describes it better (Supplementary Note~1); the parameters of all fits are listed in Supplementary Table~1. 
Three of the four frequencies require a spin--lattice term, in each case of the $\Phi_\mathrm{LRO}$ form: the magnetic contribution is confined to the ordered phase and sets in steeply below $T_\mathrm{N}$, which accounts for the anomalies of Fig.~\ref{fig:conductivity}. At $T\to0$, it amounts to
$\lambda=-9.8\pm1.4$~GHz for $\nu_1$, $+28.8\pm1.6$~GHz for $\nu_1'$ and $+102\pm9$~GHz for $\nu_2'$: the order slightly softens $\nu_1$ and stiffens $\nu_1'$ and $\nu_2'$, in each case by a few per cent at most
(Fig.~\ref{fig:parameters}a,d,h). $\nu_2$, which shows no anomaly at
$T_\mathrm{N}$, requires no spin--lattice term and decreases smoothly from 2.09~THz at 1.6~K to 1.97--1.99~THz between 160 and 191~K. The anharmonic backgrounds are also mode dependent. Their characteristic temperatures, $\Theta=310\pm27$, $184\pm14$ and $208\pm14$~K for $\nu_1$, $\nu_1'$ and $\nu_2'$, correspond to phonon frequencies $k_\mathrm{B}\Theta/h$ of 3.8--6.5~THz, within the phonon spectrum of FePS$_3$ \cite{McCreary2020,Panda2026}, but the amplitude of the softening differs strongly between the modes: $\nu_2'$ decreases by 43\% between 20 and 300~K, mostly above $T_\mathrm{N}$, a shift almost two orders of magnitude larger than the thermal shifts of the Raman-active phonons of FePS$_3$ \cite{McCreary2020}.
 
Both linewidths of mode~1 require a spin--lattice term, and both follow
$\Phi_\mathrm{SRO}$ (Fig.~\ref{fig:parameters}b). As for the frequencies, the
terms have opposite signs in the two orientations. At $\theta=0^\circ$,
$\gamma_1$ increases on cooling, from 0.24~THz at 300~K to 0.38~THz at 20~K,
opposite to the narrowing expected from anharmonic decay, with
$\eta=+0.21\pm0.07$~THz. At $\theta=90^\circ$, $\gamma_1'$ narrows on cooling,
from 0.23 to 0.16~THz, with $\eta=-0.101\pm0.010$~THz, so that the order narrows
the line well below its anharmonic background. Because $\Phi_\mathrm{SRO}$ grows
already above $T_\mathrm{N}$, both linewidths depart from their backgrounds well
above the ordering temperature. Mode~2 behaves differently
(Fig.~\ref{fig:parameters}e). $\gamma_2$ follows the anharmonic form alone,
increasing from 0.25--0.28~THz below 60~K to 0.46~THz at 191~K. $\gamma_2'$
falls from 0.33~THz at 20~K to a broad minimum between 150 and 190~K and
increases again above. It is described by the anharmonic form plus a
phenomenological step of amplitude $\eta=0.26\pm0.10$~THz, centred at
$T_c=110\pm5$~K, close to $T_\mathrm{N}$, with a width of $21\pm6$~K: the
damping of mode~2 at $\theta=90^\circ$ thus contains a contribution that
vanishes near $T_\mathrm{N}$.
 
The oscillator strengths respond to the magnetic order more strongly than the frequencies (Fig.~\ref{fig:parameters}c,f). Their spin--lattice terms follow $\Phi_\mathrm{SRO}$ for $A_1$, $A_1'$ and $A_2$, with $\zeta=0.052\pm0.015$, $0.061\pm0.010$ and $0.146\pm0.026~\Omega^{-1}\,\mathrm{cm}^{-1}\,\mathrm{THz}$,
and $\Phi_\mathrm{LRO}$ for $A_2'$, with $\zeta=-0.032\pm0.004~\Omega^{-1}\,\mathrm{cm}^{-1}\,\mathrm{THz}$. In magnitude, these terms amount to about a quarter of the low-temperature strength of $A_1$, $A_1'$ and $A_2'$, but to more than two thirds of that of $A_2$. The strength of mode~2 at $\theta=0^\circ$ is thus dominated by the term that follows the spin correlations, which accounts for its weakening above $T_\mathrm{N}$, although its
frequency shows no anomaly; phonons of FePS$_3$ whose coherent amplitudes follow the order without any frequency renormalization have likewise been reported \cite{Ergecen2023,Mertens2023}. An isolated polar phonon varies its oscillator strength in proportion to the square of the effective mode charge projected onto the THz polarization~\cite{GonzeLee1997,Baroni2001}. Mode mixing can redistribute this strength, and stacking disorder may contribute to the infrared activity of mode 1, as discussed below. For $A_1$ and $A_2$, the measured decrease is centred above $T_\mathrm{N}$: with the temperature scale of $\Phi_\mathrm{SRO}$ left free, the fits prefer $\approx$150 and $\approx$140~K, respectively, whereas $A_1'$ follows $T_\mathrm{N}$ (Supplementary Note~2). At $\theta=0^\circ$, the strengths thus respond to correlations that develop well above $T_\mathrm{N}$, which the nearest-neighbour function describes only approximately.
 
The relative weight of mode~2, $W_2^\mathrm{rel}$, summarizes how the two modes share the low-energy absorption in each orientation
(Fig.~\ref{fig:parameters}g). Below 60~K, $W_2^\mathrm{rel}\approx0.56$ at $\theta=0^\circ$ and $\approx0.37$ at $\theta=90^\circ$. On warming, the two orientations evolve in opposite directions, and by 191~K they have nearly exchanged their values, with 0.33 and 0.56, respectively. The strengths show that the mechanism differs between the orientations. At $\theta=90^\circ$, $A_1'$ decreases and $A_2'$ increases by the same amount, so that their sum
remains practically constant up to 240~K: strength is transferred from mode~1 to mode~2. At $\theta=0^\circ$, both $A_1$ and $A_2$ decrease above $\approx$90~K, $A_2$ much more strongly, and their sum falls from 0.37 to $0.24~\Omega^{-1}\,\mathrm{cm}^{-1}\,\mathrm{THz}$ between 101 and 191~K: $W_2^\mathrm{rel}$ decreases because mode~2 loses the part of its strength that is tied to the spin correlations, not because weight is transferred to mode~1.
 
The two curves cross at 118~K, at $T_\mathrm{N}$ within the temperature step, where $W_2^\mathrm{rel}$ changes fastest at $\theta=90^\circ$: the in-plane anisotropy of the weight distribution reverses its sign at the magnetic ordering, as does the splitting of mode~1 (Fig.~\ref{fig:conductivity}). We suggest that both reversals reflect two contributions of opposite sign to the in-plane anisotropy of the low-energy response. The first is the anisotropy of the paramagnetic crystal, set by the stacking. The second appears with the zigzag order, which distinguishes $a$, along which its ferromagnetic chains run, from $b$ \cite{Lancon2016}, and is accompanied by an anisotropic distortion of the lattice: $a$ and $b$ change in opposite directions below $T_\mathrm{N}$ \cite{Murayama2016}, and the stacking is sheared along $a$, with a change of the monoclinic angle of $\approx$0.1$^\circ$ that is proportional, in a Landau description, to the squared order parameter \cite{Zhou2022} and sets in almost discontinuously at $T_\mathrm{N}$ \cite{Zong2023}. If the magnetic contribution is opposite in sign to the structural one and larger at low temperature, the anisotropy reverses where the two cancel. Because the magnetic contribution grows from zero to a large fraction of its low-temperature value within a few kelvin below $T_\mathrm{N}$, as the spin--lattice terms of the frequencies do (Fig.~\ref{fig:parameters}h), the cancellation falls close to $T_\mathrm{N}$ for a wide range of ratios between the two contributions. For the splitting of mode~1, the fits make this quantitative. At $T_\mathrm{N}$, the anharmonic background of $\nu_1'$ lies 18~GHz below that of $\nu_1$, so that the reversal occurs where the spin--lattice terms have reached about half their low-temperature difference, that is, where $\Phi_\mathrm{LRO}\approx0.5$, about 1~K below $T_\mathrm{N}$; any value of $\Phi_\mathrm{LRO}$ between 0.2 and 0.7 would place the reversal within $\approx$5~K of $T_\mathrm{N}$.
 
Above $\approx$200~K, mode~2 departs from this description in both
orientations, whereas mode~1 evolves smoothly (Fig.~\ref{fig:parameters}): at $\theta=0^\circ$, $\nu_2$ and $\gamma_2$ become non-monotonic, and at $\theta=90^\circ$, $A_2'$ decreases above 240~K, so that the total strength is no longer conserved. Weaker absorption also appears above mode~2, near the upper
limit of the measured range (Fig.~\ref{fig:conductivity}a,b). We can only suggest possible origins. In the selenide analogue FePSe$_3$, paramagnon-like excitations of short-range-correlated spin clusters persist above $T_\mathrm{N}$, and the response of uncorrelated Fe$^{2+}$ ions is reached only well above it \cite{Chen2024}. As the correlations decay, magnetic spectral weight would thus move from the magnon towards the single-ion transition $|\pm2\rangle\to|\pm1\rangle$, near 2~THz \cite{Lancon2016,ZhangQ2021}, that is, into the range of mode~2. Alternatively, spin disorder, which activates acoustic phonons from the whole Brillouin zone in the Raman spectra of FePS$_3$ \cite{Sekine1990}, could add a broad absorption in the same range.
 
\subsection*{Magnetic-field response and critical field of the higher mode}\label{sec:field}
 
\begin{figure*}[t]\begin{ppfloat}
\centering
\includegraphics[width=156.8mm]{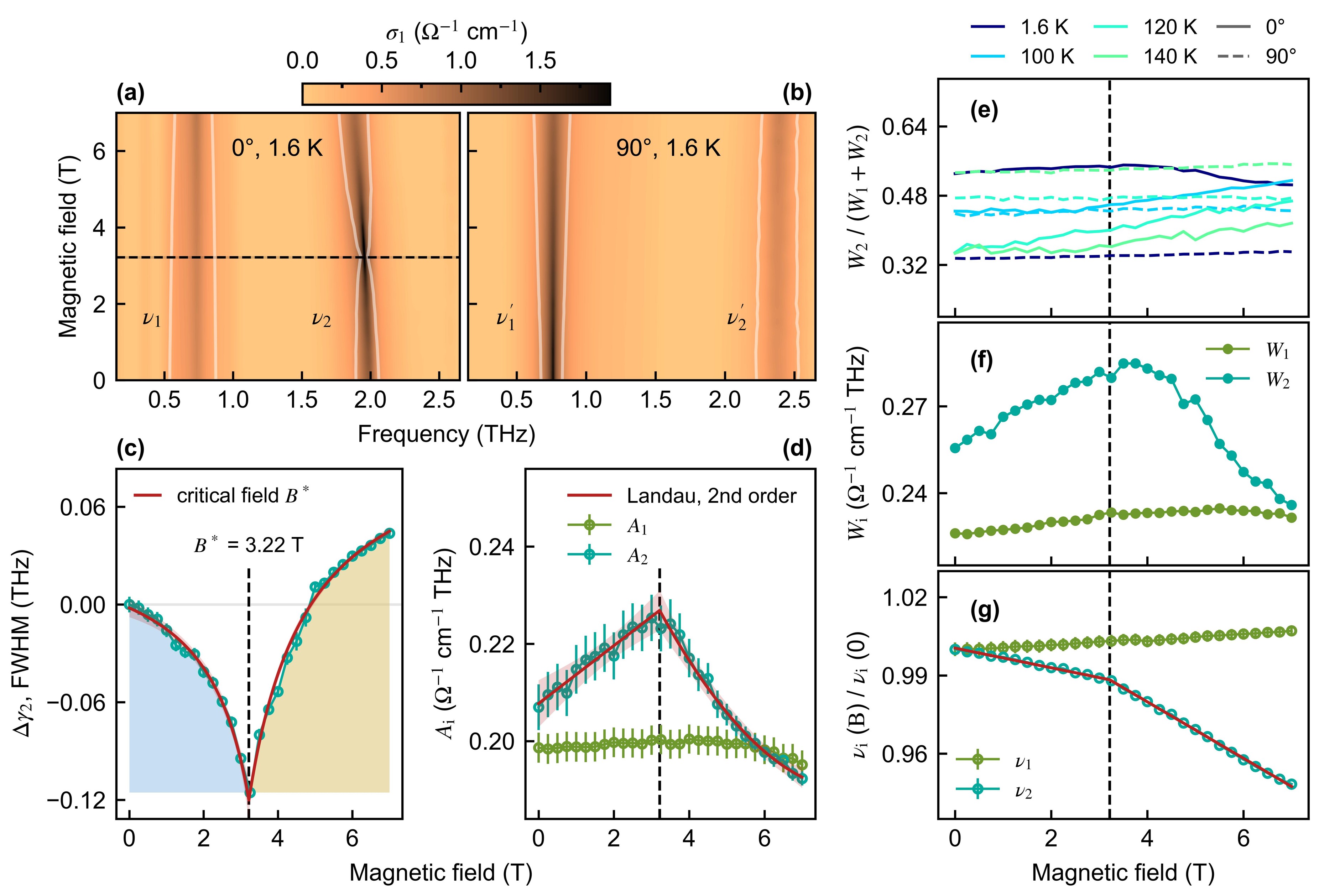}
\caption{\textbf{Magnetic-field dependence at 1.6~K and critical field of the higher mode at $\theta=0^\circ$.} \textbf{a},~\textbf{b} Colour maps of the baseline-subtracted conductivity, $|\sigma_1-\sigma_b|$, as a function of frequency and field at 1.6~K for $\theta=0^\circ$ and $\theta=90^\circ$, with $\mathbf{B}\parallel c^*$, on a common colour scale. Light lines mark $\nu_i\pm\gamma_i/2$ from the fits (for mode~2 at $\theta=0^\circ$, with $\gamma_2$ defined below), and the dashed line in \textbf{a} marks $B^*$. \textbf{c} Change of the linewidth (FWHM) of mode~2 with respect to zero field, $\Delta\gamma_2=\gamma_2(B)-\gamma_2(0)$, \textbf{d} oscillator strengths $A_1$ and $A_2$ and \textbf{g} normalized frequencies $\nu_i(B)/\nu_i(0)$ at $\theta=0^\circ$ and 1.6~K. Where mode~2 is described by two components with a common centre (0--4.75~T; Methods), $\gamma_2$ is the FWHM of their sum and $A_2$ the sum of their strengths. Solid red lines in \textbf{c} and \textbf{g} are a joint fit of $\nu_2$ and $\gamma_2$ with a common critical field, $B^*=3.217\pm0.024$~T (Supplementary Note~3): $\nu_2$ is continuous and changes slope at $B^*$, where $\gamma_2$ has a cusp-like minimum. Shaded areas in \textbf{c} mark $\Delta\gamma_2$ above its minimum, on the low-field (blue) and high-field (beige) sides. The solid red line in \textbf{d} is a Landau expansion of $A_2$ to second order in $(B-B^*)$, with $B^*$ fixed at the same value; $A_1$ is shown without a fit. Light red areas are 95\% confidence bands, and error bars are the uncertainties of the fits. \textbf{e} Relative spectral weight of the higher mode, $W_2/(W_1+W_2)$, at 1.6, 100, 120 and 140~K for $\theta=0^\circ$ (solid lines) and $\theta=90^\circ$ (dashed lines). \textbf{f} Spectral weights $W_1$ and $W_2$ at $\theta=0^\circ$ and 1.6~K. The weights are integrals of $|\sigma_1-\sigma_b|$ minus the fitted oscillator of the other mode, over windows set for each spectrum (Methods). Dashed vertical lines mark $B^*$.}\label{fig:field}
\end{ppfloat}\end{figure*}

Figure~\ref{fig:field} shows the response to a magnetic field applied normal to the layers. At 1.6~K, both modes can be followed continuously up to 7~T in both orientations, with no resolved splitting and no additional absorption (Fig.~\ref{fig:field}a,b). The response is strongly selective. At $\theta=90^\circ$, neither mode changes appreciably, and the relative weight of mode~2 remains practically constant at all four temperatures measured (Fig.~\ref{fig:field}e). At $\theta=0^\circ$, the frequency and strength of mode~1 are also practically field independent (Fig.~\ref{fig:field}d,g), whereas $\gamma_1$ narrows by $\approx$0.07~THz up to 7~T (Supplementary Table~3). The field acts mainly on mode~2 at $\theta=0^\circ$.
 
Relative to zero field, $\nu_2$ softens by 0.10~THz at 7~T
(Fig.~\ref{fig:field}g), and the softening accelerates above $\approx$3~T. Near the same field, $\gamma_2$ passes through a sharp, cusp-like minimum (Fig.~\ref{fig:field}c): it narrows by $\approx$0.12~THz between zero field and $\approx$3.2~T and broadens again at higher fields, exceeding its zero-field value by $\approx$0.04~THz at 7~T. Near the minimum, the width of the summed
line is set by its narrower component and should not be read as the damping of a single oscillator. $A_2$ increases slightly up to $\approx$3~T and decreases at higher fields (Fig.~\ref{fig:field}d), whereas $A_1$ shows no anomaly. The three quantities change their behaviour at a single field, $B^*=3.217\pm0.024$~T (Fig.~\ref{fig:field}c,g; Supplementary Note~3): at $B^*$, the softening rate of $\nu_2$ increases nearly threefold, and around it $\gamma_2$ increases linearly with $|B-B^*|$ on both sides of the minimum. Separate fits of $\nu_2$, $\gamma_2$ and $A_2$ give compatible values of the critical field.
 
The spectral weights show the same selectivity (Fig.~\ref{fig:field}e,f). At $\theta=0^\circ$ and 1.6~K, $W_2$ increases up to 3.5--3.75~T and decreases above, while $W_1$ remains practically constant, so that $W_2^\mathrm{rel}$ has a shallow maximum near 3.5~T. At 100~K, $W_2^\mathrm{rel}$ is constant up to $\approx$2.5~T and increases above, whereas at 120 and 140~K it increases from zero field. The field scale is thus still visible at 100~K, below $T_\mathrm{N}$, but not at 120 and 140~K, above it, which ties it to the ordered phase.
 
$B^*$ is about ten times lower than the fields of the transitions reported for FePS$_3$ with $\mathbf{B}\parallel c^*$, the spin flip near 35~T~\cite{ZhangQ2021} and the magnetization steps at 39 and 41~T~\cite{Wildes2020}, and the static magnetization, measured on another crystal, shows no transition between 0 and 5~T~\cite{Wildes2020}. The softening of $\nu_2$ is also large on the scale of the field-induced phonon shifts reported for FePS$_3$: at 7~T, it is fifteen times the largest of them, found for a phonon 0.4~THz below the magnon~\cite{Mertens2023}, and exceeds by two orders of magnitude the magnon-mediated shift expected for a mode near 2~THz from the magnon--phonon couplings of FePS$_3$~\cite{Vaclavkova2021,ZhangQ2021,Liu2021}. Neither a smooth crossover between two field-dependent relaxation channels nor a resonance with an excitation crossing the mode reproduces the cusp of $\gamma_2$ (Supplementary Note~3). 
A simultaneous change of frequency, linewidth and strength at one field is the behaviour expected when a field-induced order parameter couples to the mode. In Supplementary Note~3, we show that a Landau expansion in this order parameter, with relaxational dynamics \cite{HohenbergHalperin1977}, describes $\nu_2$, $\gamma_2$ and $A_2$ jointly. The departure of $\nu_2$ from its low-field trend grows as $(B-B^*)^p$ with $p=0.97\pm0.02$, the mean-field value, which excludes
the exponents of the two- and three-dimensional Ising models (1/4 and
$\approx$0.65). The minimum of $\gamma_2$ is the counterpart, for a terahertz mode, of critical slowing down: when the relaxation time of the order parameter is long compared with the period of the mode, the damping contributed by the coupling is proportional to its relaxation rate, which vanishes at $B^*$; the more slowly the order parameter relaxes, the less energy it can absorb from the oscillation. A Landau free energy with relaxational dynamics has been used to describe the critical slowing down of the interlayer shear of FePS$_3$ near $T_\mathrm{N}$ \cite{Zhou2022}; here, the control parameter is the field, and the probe a terahertz phonon. Within the effective-oscillator description, the changes in slope of the total band strength $A_2$ and the resonance frequency $\nu_2$ at $B^*$ imply that the fractional response of the projected mode effective charge to $\langle\psi^2\rangle$ is about two to three times that of the effective force constant (Supplementary Note 3). The microscopic identity of the order parameter is not determined by our data. Each zigzag layer carries no net moment along $c^*$ \cite{Lancon2016}, so that a
field normal to the layers couples to the order only at second order; together with the absence of a transition in the static magnetization, this leads us to identify $B^*$ as the field at which the response of mode~2 changes, without assigning it to a thermodynamic phase transition.
 
\subsection*{Origin of the low-energy resonances}\label{sec:origin}
 
The four resonances are not magnetic excitations. Besides the excitations of the ordered phase, excluded by the persistence of the resonances far above $T_\mathrm{N}$, no magnetic mode has been found in FePS$_3$ below $\approx$15~meV \cite{Lancon2016,McCreary2020,Wyzula2022,Dhakal2024}. Transitions between the single-ion levels of Fe$^{2+}$ split by the anisotropy fall in the same frequency range, at $\Delta\approx0.64$--0.71~THz and $3\Delta\approx1.93$--2.12~THz \cite{Lancon2016,Liu2021}, but the temperature and field dependences exclude them. The strength of a transition between localized levels scales with their population difference, which at 301~K would leave a transition at the frequency of mode~1 with 7\% of its 20~K strength. Instead, the strength of mode~1 summed over the two orientations, $A_1+A_1'$, is unchanged up to 101~K and still retains 70\% of its 20~K value at 301~K, and $A_2'$ even increases on warming. In the ordered phase, moreover, the exchange field shifts the transition from the ground $|\pm2\rangle$ levels to $|\pm1\rangle$ up to the magnon frequency \cite{Lancon2016} and leaves the $|\pm1\rangle$ levels, the initial states of the $\Delta$ transition, empty at 1.6~K. Finally, a transition with $\Delta S_z=\pm1$ would shift by $\approx$0.2~THz at 7~T for $g\approx2$, whereas $\nu_1$ shifts by only 5~GHz (Fig.~\ref{fig:field}g).
 
The lower pair is consistent with an interlayer shear resonance, with $\nu_1$ and $\nu_1'$ as its two in-plane components. In the ideal $C2/m$ crystal, the primitive cell contains a single layer \cite{Murayama2016}, so that the interlayer shear and breathing motions belong to the acoustic branches and have no infrared-active counterpart at the zone centre. This distinguishes FePS$_3$ from crystals that stack inequivalent layers, such as the vdW mineral clinochlore, in which rigid-layer shear modes between talc-like and brucite-like layers are infrared active~\cite{kawahala_shaping_2025}. Bulk FePS$_3$, however, contains rotational twins and stacking faults located at the vdW gaps \cite{Murayama2016,Lancon2016,Wildes2020}, favoured by the nearly
hexagonal in-plane metric of the monoclinic cell~\cite{Murayama2016}. 

Because each layer is centrosymmetric and nearly trigonal, two adjacent layers are related by an inversion centre whatever their registry, and their relative displacement alone carries no dipole. At a fault, however, the layer whose two neighbours are shifted along different directions loses its inversion centre, and its displacement relative to both neighbours can carry a dipole. Being planar, the faults relax only the wavevector component normal to the layers: light then couples to the shear branch at zero in-plane wavevector, with a weight that vanishes at the zone centre and is largest at the zone boundary, where the density of states is also singular. The induced absorption is therefore concentrated near, but not above, the zone-boundary frequency, and $K=\mu_A(\pi\nu)^2$ (Fig.~\ref{fig:parameters}i) is a lower bound to the interlayer force constant per unit area. At low temperature, $\nu_1$ and $\nu_1'$
give lower bounds of 1.0 and $1.1\times10^{19}$~N\,m$^{-3}$, comparable to the shear force constant of the isostructural NiPS$_3$, $1.42\times10^{19}$~N\,m$^{-3}$~\cite{Nguyen2025}, and of other layered
crystals~\cite{Tan2012,ZhangX2013}. The oscillator strength of the pair is nevertheless large for a dipole that vanishes for any pair of adjacent layers and arises only at the faults; local vibrations of point defects, such as weakly bound species in the vdW gaps~\cite{BarkerSievers1975}, could also contribute, although their frequencies would not be tied to the interlayer stiffness.
 
The shear stiffness of FePS$_3$ itself has been obtained from the velocity of its interlayer-shear acoustic wave. Ultrafast x-ray diffraction on $\approx$200-nm-thick flakes gives $1668\pm18$~m\,s$^{-1}$ from the dispersion and $1709\pm15$~m\,s$^{-1}$ from the oscillation period, in agreement with the 1715~m\,s$^{-1}$ calculated for shear along $a$~\cite{Zhou2026}. In the chain model of Fig.~\ref{fig:parameters}i, with an interlayer spacing of 6.42~\AA~\cite{Murayama2016,Wang2016}, these velocities correspond to $K=1.34$--$1.42\times10^{19}$~N\,m$^{-3}$ and to a zone-boundary frequency of 0.83--0.85~THz. Between 20 and 101~K, $\nu_1$ and $\nu_1'$ lie 8--16\% below this frequency. Part of this difference is expected, since the fault-induced absorption lies near, but not above, the zone-boundary frequency. A weak force constant between next-nearest layers, $K_2$, may account for the rest: it raises the sound velocity, which depends on $K+4K_2$, relative to the zone-boundary frequency, which depends on $K$ alone.
Attributed entirely to $K_2$, the difference would give $K_2/K\approx0.05$--0.11, which is therefore an upper bound. The measured frequencies are also close to the $\approx$0.74~THz implied by the velocity calculated for shear along $b$~\cite{Zhou2026}. A lower velocity, $1120\pm20$~m\,s$^{-1}$, measured in free-standing films 16--57~nm thick~\cite{Zong2023}, would place the zone-boundary frequency at 0.56~THz, below $\nu_1$ and $\nu_1'$, frequencies that the fault-induced absorption could then not reach. 

Zhou et~al.\ attribute the difference between the two measurements to the smaller thickness of those films, through surface oxidation, and to their magnetic state after photoexcitation \cite{Zhou2026}; we therefore take the velocity of the thicker flakes, reproduced by the calculation, as the reference for the bulk crystal. The same rigid-layer model provides a comparison with interlayer breathing. Using the room-temperature longitudinal sound velocities reported for FePS$_3$~\cite{Zhou2026} gives a zone-boundary frequency of 1.77–2.08~THz, well above the 0.72 and 0.76 THz resonances measured at 20~K. The proximity of both resonances to the transverse acoustic scale, together with the independently established coupling of interlayer shear to magnetic order, favours an interlayer shear origin for the lower pair. Within this reading, the anomalies of $\nu_1$ and $\nu_1'$ at $T_\mathrm{N}$ measure the coupling between the interlayer shear and the magnetic order. The anharmonic backgrounds of the two components extrapolate to the same frequency at $T\to0$, 0.73~THz, so that their splitting at 20~K, 39~GHz, is entirely magnetic: $\lambda'-\lambda=38.6\pm2.1$~GHz. The zigzag order thus stiffens the shear along one in-plane direction and softens it along the other, and its abrupt onset fixes the reversal of the splitting at $T_\mathrm{N}$. Above $T_\mathrm{N}$, the two components differ mainly in frequency: at 300~K, their linewidths and strengths are nearly equal. This coupling of the interlayer shear to the zigzag order agrees with the spin--shear coupling identified in time-resolved experiments~\cite{Zhou2022,Zong2023,Zhou2026}. Because rotational twins mix the two in-plane responses \cite{Lancon2016,Wildes2020}, the measured differences between $\theta=0^\circ$ and $90^\circ$, including $\lambda'-\lambda$, are lower bounds, in magnitude, to the intrinsic ones.
 
Mode~2 behaves differently. $\nu_2$ and $\nu_2'$ have different anharmonic backgrounds, cross near 150~K (Fig.~\ref{fig:parameters}d) and respond differently to the magnetic order and to the field; they are therefore better regarded as distinct lattice modes than as two components of one mode. $\nu_2'$ behaves as a strongly anharmonic polar phonon, whose coupling to the magnetic order is small compared with its anharmonic renormalization. $\nu_2$, whose frequency shows no anomaly at $T_\mathrm{N}$, carries a strength largely tied to the spin correlations and the response at $B^*$; its large field-induced softening points to a much stronger magnetoelastic coupling, whose microscopic origin remains open. Mode~2 also lies in a range covered by absolute infrared spectra of FePS$_3$, which do not report it \cite{JouanneJulien1988,Panda2026}. For the 26-$\mu$m-thick slab of Panda et al. \cite{Panda2026}, an absorption with the strength measured here would reduce the transmission by 17--25\% near 66--72~cm$^{-1}$, about ten times the noise of those spectra (Supplementary Note~4), so that the smaller thickness does not explain its absence; conversely, the broad absorption near 52~cm$^{-1}$ reported in that work is several times weaker in our crystal.

For $\nu_2$, our reanalysis of the bulk force data of Haddadi et al.~\cite{Haddadi2024,HaddadiData2024} identifies a vibration near $68.2~\mathrm{cm}^{-1}$ (2.04~THz) that bends the intralayer Fe--S--Fe bridges. This motion could modulate the exchange between neighbouring Fe sites through the bridging S atoms~\cite{Pestka2024,Wang2025FePX3}. It thus provides a possible microscopic basis for the spin--correlation dependence of the fitted oscillator strength and the field response near $B^*$. For $\nu_2'$, a distinct Fe-dominated vibration at $79$--$80~\mathrm{cm}^{-1}$ lies close to the low-temperature resonance and produces much less bridge bending. This second assignment rests mainly on the frequency match at low temperature. Whether the same vibration dominates the band as it softens to 1.35~THz at 300~K remains open. Both candidates are even under inversion in the ideal structure, so their contribution to electric-dipole absorption requires relaxation of the ideal selection rules (Supplementary Note~6).

These differences suggest that the infrared strength of $\nu_2$ may depend on local structure. Rotational twins have been reported in bulk FePS$_3$~\cite{Lancon2016,Murayama2016}, with boundaries
observed in the van der Waals gaps~\cite{Murayama2016}. For the $68.2~\mathrm{cm}^{-1}$ candidate, the distortion of the Fe--S--Fe bridges changes sign under an in-plane translation. If disorder near a twin boundary breaks this translation and locally lifts inversion symmetry, the contributions from neighbouring bridges could acquire a net dynamic dipole along the THz polarization. The comparison between crystals is consistent with this mechanism, although it does not identify the defect responsible.
 
Taken together, the data reveal two channels through which the magnetic order acts on the low-energy lattice response of FePS$_3$. The frequencies that respond to the order follow the long-range order and change only below $T_\mathrm{N}$, whereas the linewidths of mode~1 and three of the four oscillator strengths follow the short-range correlations and change over a wider range, extending above $T_\mathrm{N}$. Both channels are selective with respect to the mode and to the in-plane polarization, to the point of reversing the in-plane anisotropy at $T_\mathrm{N}$, and the same selectivity holds in field, where only mode~2 at $\theta=0^\circ$ responds at $B^*$.

\section*{Methods}
\subsection*{Synthesis of FePS$_3$ crystals}
FePS$_3$ crystals were prepared by chemical vapour transport (CVT) using iodine
as a transport agent in a sealed quartz ampoule. Stoichiometric amounts of iron powder (99.9\%, $-100$ mesh, Strem, USA), sulfur granules (99.9999\%, 2--6~mm, Wuhan Xinrong New Materials Co., China) and phosphorus granules (99.9999\%, 2--6~mm, Wuhan Xinrong New Materials Co., China), corresponding to a target yield of 20~g of FePS$_3$, were placed in a quartz ampoule ($40\times250$~mm). Iodine (0.8~g, 99.9\%, Fisher Scientific, UK) was added as
the transport agent. The ampoule was flame-sealed under high vacuum below $1\times10^{-3}$~Pa using an oil diffusion pump, a liquid-nitrogen cold trap and an oxygen--hydrogen welding torch. The sealed ampoule was first placed in a muffle furnace and gradually heated to 600~$^\circ$C using the following temperature programme: 450~$^\circ$C for 25~h, 500~$^\circ$C for 50~h and 600~$^\circ$C for 50~h. The heating and cooling rates were 1~$^\circ$C\,min$^{-1}$. Subsequently, the ampoule was transferred to a
horizontal two-zone furnace for CVT crystal growth. In the first step, the
growth zone was heated to 750~$^\circ$C while the source zone was maintained at
600~$^\circ$C for 2 days. The thermal gradient was then reversed, and the growth
zone was maintained at 650~$^\circ$C while the source zone was heated to 750~$^\circ$C for 14 days. Finally, the ampoule was cooled to room temperature and opened in an argon-filled glovebox.
 
\subsection*{Measurements}
The temperature and magnetic-field dependence of the terahertz response was investigated by transmission terahertz time-domain spectroscopy (THz-TDS) between 1.6 and 300~K and in applied fields up to 7~T. The measurements employed a fiber-coupled photoconductive-antenna spectrometer (TeraFlash Pro, TOPTICA Photonics) integrated with a cryofree superconducting magnet (SpectromagPT, Oxford Instruments). The sample was mounted over an aperture of 2.5~mm diameter, with off-axis parabolic mirrors focusing the incident beam onto the sample and collecting the transmitted radiation~\cite{Dias2026}. The THz propagation direction was parallel to the applied magnetic field, corresponding to Faraday geometry. To probe the in-plane anisotropy, spectra were acquired for two sample orientations, labelled $\theta=0^\circ$ and $90^\circ$, related by a $90^\circ$ rotation of the sample within its plane relative to the fixed polarization of the incident THz electric field. At each temperature, a reference waveform was recorded without the sample at $B_{\mathrm{ext}}=0$~T. Fourier transformation of the sample and reference waveforms yielded the frequency-domain electric fields $E_{\mathrm{sam}}(\nu)$ and $E_{\mathrm{ref}}(\nu)$, from which the complex transmission was calculated as $T(\nu)=E_{\mathrm{sam}}(\nu)/E_{\mathrm{ref}}(\nu)$. The signal-processing procedure followed Ref.~\cite{Marulanda2025}. The complex optical conductivity, $\sigma(\nu)=\sigma_1(\nu)+i\sigma_2(\nu)$, was subsequently retrieved by inverting the Fresnel transmission of a plane-parallel slab~\cite{Lloyd-Hughes2012-es}. These spectra were used to determine the resonance frequencies, oscillator strengths, and effective scattering times.
 
\subsection*{Oscillator fits}
Each resonance was described by a damped harmonic oscillator,
\begin{equation}\label{eq:dho}
\sigma_\mathrm{DHO}(\nu)=\frac{-iA\nu}{\nu_0^2-\nu^2-i\gamma\nu},
\end{equation}
whose real part peaks at the resonance frequency $\nu_0$, has a full width at
half maximum $\gamma$ and integrates to $\pi A/2$, with $A$ the oscillator
strength. The real and imaginary parts of the conductivity were fitted
simultaneously, with $\nu_0$, $\gamma$ and $A$ common to both. The non-resonant
background of $\sigma_1$ was described by a smooth baseline, $\sigma_b$,
obtained by asymmetric least-squares smoothing \cite{EilersBoelens2005}, and
that of $\sigma_2$ by a linear term. The uncertainties are the standard errors of the fits, and the fits of all spectra are shown in Supplementary Figs.~1 and 2. In field, mode~2 at $\theta=0^\circ$ was described by two oscillators with a common centre, one narrow and one broad, from 0 to 4.75~T, and by a single oscillator from 5 to 7~T (Supplementary Note~5). The integration windows of the spectral weights in Fig.~\ref{fig:field}e,f
extended, on each side of a line, to where the integrand reached the noise
level.
 
\subsection*{Temperature dependence of the mode parameters}
Each parameter in Fig.~\ref{fig:parameters}a--f was described by an anharmonic
background plus a spin--lattice term proportional to $\Phi(T)$, a spin
correlation function normalized to $\Phi(0)=1$:
\begin{align}
\nu(T)&=\left[a+b\coth\left(\frac{\Theta}{2T}\right)\right]^{1/2}+\lambda\Phi(T),
\label{eq:nu}\\
\gamma(T)&=\gamma_0+c\coth\left(\frac{\Theta_\gamma}{2T}\right)+\eta\Phi(T),
\label{eq:gamma}\\
A(T)&=A_0+c_AT+\zeta\Phi(T).\label{eq:A}
\end{align}
The first terms are the anharmonic backgrounds~\cite{Barrett1952,Klemens1966,Balkanski1983,MenendezCardona1984},
with characteristic temperatures $\Theta$ and $\Theta_\gamma$. $\Phi$ is either
the squared order parameter, $\Phi_\mathrm{LRO}=[M(T)/M(0)]^2$, or the
nearest-neighbour spin correlation, $\Phi_\mathrm{SRO}=\langle\sigma_i\sigma_j\rangle$,
of the Ising model on the honeycomb lattice, both evaluated from its exact
solution~\cite{Houtappel1950,Potts1952,Baxter1982} with the transition at
$T_\mathrm{N}=118$~K. For each parameter, the spin--lattice term and the form
of $\Phi$ were selected by statistical model comparison (Supplementary Note~1). For $\gamma_2'$, the spin--lattice term was replaced by a phenomenological
step, $\eta S(T)$, with $S(T)=[1+e^{(T-T_c)/w}]^{-1}$. Each model was fitted
only at the temperatures at which the corresponding parameter is well determined
(ranges in Supplementary Table~1); the other points are shown in Fig.~\ref{fig:parameters} but were not used in
the fits. For the spectra at $\theta=90^\circ$ in which the transmission at the
centre of mode~2 reaches the noise floor (161--176~K), a systematic uncertainty,
estimated from the values at neighbouring temperatures, was added to that of the
fit. The 95\% confidence bands follow from the full covariance of the fitted
parameters. The critical-field model used for mode~2 in
Fig.~\ref{fig:field}c,d,g is described in Supplementary Note~3. 
\subsection*{Linear-chain model}
For a linear chain of rigid layers with mass per unit area $\mu_A$, spacing $d$
and force constant per unit area $K$ between adjacent layers, the shear branch
has the zone-boundary frequency $\nu_\mathrm{ZB}=(1/\pi)(K/\mu_A)^{1/2}$ and the
sound velocity $v=d(K/\mu_A)^{1/2}$. A force constant $K_2$ between next-nearest
layers leaves $\nu_\mathrm{ZB}$ unchanged and raises the sound velocity to
$v=d[(K+4K_2)/\mu_A]^{1/2}$. For FePS$_3$, $\mu_A=1.985\times10^{-6}$~kg\,m$^{-2}$
is the mass of the four formula units of the monoclinic cell divided by its $ab$
area, and $d=c\sin\beta=6.42$~\AA, both from the lattice
parameters \cite{Murayama2016}.

\begin{ppendmatter}
\textbf{Data Availability:} All data generated or analyzed during this study are included in this published article and its supplementary information file.

\textbf{Acknowledgments:} Activities within the GCTI-THz/USP group were supported by the São Paulo Research Foundation (FAPESP), Grants Nos. 2021/12470-8, 2023/16742-8 and 2026/15952-7. E.D.S. acknowledges support from FAPESP Grant No. 2025/02029-3. I.D.B. and A.C.F.B. acknowledge financial support from FAPESP through the Research, Innovation and Dissemination Center for Molecular Engineering for Advanced Materials – CEMol (Grant CEPID Nos. 2024/00989-7 and 2025/27155-1). Z.S. was supported by ERC-CZ program (project LL2101) from Ministry of Education Youth and Sports (MEYS). F.G.G.H. acknowledges financial support from Grant No. 306550/2023-7 of the National Council for Scientific and Technological Development (CNPq). All authors thank the Brazilian Synchrotron Light Laboratory (LNLS) and Brazilian Nanotechnology National Laboratory (LNNano), part of the Brazilian Centre for Research in Energy and Materials (CNPEM), a private non-profit organization under the supervision of the Brazilian Ministry for Science, Technology, and Innovations (MCTI), for sample preparation (LAM proposal 20253692) and Raman spectroscopy measurements (LNNano 20250171).

\textbf{Author contributions:} F.G.G.H. and I.D.B. conceived the project. J.P. and Z.S. grew the samples. A.C.F.B. and B.R.A.N. contributed to sample preparation and preliminary characterization. N.M.K. built the magneto-terahertz spectrometer and assisted E.D.S. with the measurements. E.D.S. analyzed the experimental data, prepared the figures and wrote the manuscript with input from F.G.G.H. F.G.G.H. supervised the project. All authors read and approved the final manuscript.

\textbf{Competing Interests:} All authors declare no financial or non-financial competing interests.
\end{ppendmatter}

\printbibliography

\onecolumn
\nobalance
\begin{refsection}
\begin{ppsupplement}
\begin{center}
{\Large\bfseries Supplementary Information}\\[6pt]
{\large Two channels of magnetic renormalization in the terahertz lattice response of FePS$_{3}$}\\[4pt]
{\normalsize Eduardo D. Stefanato$^{1}$, Nicolas M. Kawahala$^1$, Ana Carolina F. de Brito$^2$, \\
Jan Plutnar$^3$, Zdenek Sofer$^3$, Bernardo R. A. Neves$^4$,\\
Ingrid D. Barcelos$^2$ and Felix G. G. Hernandez$^{1,\dagger}$\\}
\end{center}

\bigskip
\noindent\textbf{Contents}

\smallskip
\noindent
Supplementary Note~1: Temperature models: spin correlation functions and model selection\\
Supplementary Note~2: Temperature scale of the spin correlations in the oscillator strengths\\
Supplementary Note~3: Critical-field model of mode~2 at $\theta=0^\circ$\\
Supplementary Note~4: Mode~2 in published absorption spectra of thinner samples\\
Supplementary Note~5: Oscillator fits and two-component description of mode~2 in field\\
Supplementary Note~6: Candidate lattice vibrations for the upper THz bands\\
Supplementary Figs.~1 and 2: Oscillator fits of the temperature and field series\\
Supplementary Tables~1--6

\snote{Temperature models: spin correlation functions and model selection}\label{sn:selection}

\paragraph{Spin correlation functions.}
The spin--lattice terms of Eqs.~(2)--(4) of Methods are proportional to a spin
correlation function, $\Phi(T)$, of the two-dimensional Ising model with
nearest-neighbour coupling on the honeycomb lattice, the geometry of the Fe
sublattice. The temperature enters through the reduced coupling
$K=K_\mathrm{c}T_\mathrm{N}/T$, with $K_\mathrm{c}=\frac12\ln(2+\sqrt3)\approx0.6585$
and $T_\mathrm{N}=118$~K. Decimation of one sublattice maps the honeycomb lattice
onto a triangular lattice with coupling $K'$, where
$e^{4K'}=2\cosh2K-1$ (star--triangle relation~\cite{Baxter1982}). With the
spontaneous magnetization of the triangular lattice~\cite{Potts1952}, the squared
order parameter is
\begin{equation}
\Phi_\mathrm{LRO}=M^2,\qquad M^8=1-\frac{16u^3}{(1+3u)(1-u)^3},\qquad
u=\frac{1}{2\cosh2K-1},
\label{eq:phi_lro}
\end{equation}
for $T<T_\mathrm{N}$, and $\Phi_\mathrm{LRO}=0$ above. The nearest-neighbour
correlation follows from the free energy per site of the honeycomb
lattice~\cite{Houtappel1950},
\begin{gather}
\Phi_\mathrm{SRO}=\langle\sigma_i\sigma_j\rangle=\frac23\,\frac{\partial(\ln Z/N)}{\partial K},
\label{eq:phi_sro}\\
\frac{\ln Z}{N}=\frac34\ln2+\frac{1}{16\pi^2}\int_0^{2\pi}\!\!\int_0^{2\pi}
\ln\left[c^3+1-s^2\left(\cos\omega_1+\cos\omega_2+\cos(\omega_1+\omega_2)\right)\right]
d\omega_1\,d\omega_2,
\label{eq:houtappel}
\end{gather}
with $c=\cosh2K$ and $s=\sinh2K$; the derivative can be written in closed form
with the complete elliptic integral of the first kind. $\Phi_\mathrm{SRO}$ equals
$4/(3\sqrt3)\approx0.770$ at $T_\mathrm{N}$, approaches $\tanh K$ well above it
and is still 0.26 at 300~K. For the test of robustness below, we also used the
functions of the square lattice, with $K_\mathrm{c}=\frac12\ln(1+\sqrt2)$:
$\Phi_\mathrm{LRO}=(1-\sinh^{-4}2K)^{1/4}$ below
$T_\mathrm{N}$~\cite{Yang1952}, and
$\Phi_\mathrm{SRO}=\frac12\coth2K\,[1+\frac2\pi(2\tanh^22K-1)\,\mathcal{K}(\kappa)]$,
with $\kappa=2\sinh2K/\cosh^22K$ and $\mathcal{K}$ the complete elliptic integral
of the first kind~\cite{Onsager1944}.

\paragraph{Fits and selection criteria.}
For each parameter of Fig.~3a--f of the main text, three models were fitted: the
anharmonic background alone, and the background plus a term proportional to
$\Phi_\mathrm{LRO}$ or to $\Phi_\mathrm{SRO}$. Except for $\gamma_2'$ and $A_2'$,
the fits are unweighted least-squares fits, and the uncertainties of the
parameters follow from their covariance matrix, scaled by the residual variance.
The characteristic temperatures $\Theta$ and $\Theta_\gamma$ were allowed to vary
between 5 and 500~K, starting from several initial values, and $\gamma_0$ and $c$
were constrained to be non-negative. The models were compared by the Bayesian
information criterion~\cite{Schwarz1978},
\begin{equation}
\mathrm{BIC}=N\ln(\mathrm{RSS}/N)+p\ln N,
\label{eq:bic}
\end{equation}
where $N$ is the number of points, RSS the residual sum of squares and $p$ the
number of parameters. A spin--lattice term was retained only if it lowered the
BIC by at least 6, which corresponds to strong evidence on the scale of Kass and
Raftery~\cite{KassRaftery1995}, and of the two forms of $\Phi$ the one with the
lower BIC was kept.

The fits of $\gamma_2'$ and $A_2'$ are weighted by the total uncertainties of the
points. For the three spectra at $\theta=90^\circ$ in which the transmission at
the centre of mode~2 reaches the noise floor (161, 171 and 176~K), the total
uncertainty adds in quadrature, to the uncertainty of the oscillator fit, a
systematic contribution, taken as the deviation of each parameter from a parabola
through its values at the two neighbouring temperatures on each side (141, 151,
191 and 200~K). In the weighted fits, $\mathrm{BIC}=\chi^2+p\ln N$, and the
uncertainties of the parameters are scaled by $\max(\chi^2_\nu,1)^{1/2}$, where
$\chi^2_\nu$ is the reduced $\chi^2$. For $\gamma_2'$, neither form of $\Phi$
describes the data ($\chi^2_\nu=98$ and 140 with $\Phi_\mathrm{LRO}$ and
$\Phi_\mathrm{SRO}$, against 503 for the background alone), whereas the background
plus the phenomenological step of Methods gives $\chi^2_\nu=15$. The 95\%
confidence bands of Fig.~3 were obtained by the delta method from the full
covariance matrix of the parameters, with the quantile of Student's $t$
distribution for $N-p$ degrees of freedom.

\paragraph{Results.}
Supplementary Table~\ref{tab:fig3} lists the parameters of all fits, with the
temperature ranges and the BIC differences. All parameters require a
spin--lattice term except $\nu_2$ and $\gamma_2$, for which the better form of
$\Phi$ changes the BIC by $-1.7$ and $+0.8$. Where the term is retained, the two
forms of $\Phi$ are clearly distinguished, by BIC differences of 11--42 (564 in
the weighted fit of $A_2'$), except for $A_1'$, for which $\Phi_\mathrm{SRO}$ is
preferred by 4.3. The term of $A_1$ passes the criterion with
$\Delta\mathrm{BIC}=7.2$, close to the threshold; the limits of this fit are
discussed in Supplementary Note~\ref{sn:scale}.

\paragraph{Robustness to the lattice.}
Replacing the honeycomb-lattice functions by those of the square lattice changes
neither the selected form of $\Phi$ nor the sign of any spin--lattice term
(Supplementary Table~\ref{tab:lattice}). The amplitudes of the terms proportional
to $\Phi_\mathrm{LRO}$ change by less than 4\%, and those of the terms
proportional to $\Phi_\mathrm{SRO}$ by 19--27\% (54\% for the weakly constrained
term of $A_1$), within 1.6 standard deviations; the difference comes from the
tail of $\Phi_\mathrm{SRO}$ above $T_\mathrm{N}$, which is larger on the
honeycomb lattice. With the square-lattice function, the term of $A_1$ falls below
the selection threshold ($\Delta\mathrm{BIC}=2.7$). The fitted curves obtained with
the two lattices differ by at most 4.2\% of the range of the data.

\begin{table}[p]
\caption{\textbf{Parameters of the temperature models of Fig.~3.} Fits of
Eqs.~(2)--(4) of Methods, with $\Phi$ from the Ising model on the honeycomb
lattice and $T_\mathrm{N}=118$~K (Supplementary Note~\ref{sn:selection}). Primes
denote $\theta=90^\circ$. Range and $N$: temperatures and number of the points
included in each fit. $\Delta\mathrm{BIC}_\mathrm{SL}$: decrease of the BIC
produced by the spin--lattice term (for $\nu_2$ and $\gamma_2$, by the better form
of $\Phi$, which is not retained). $\Delta\mathrm{BIC}_\Phi$: BIC difference between
the two forms of $\Phi$, in favour of the one retained. rms: root-mean-square
residual. Uncertainties, in parentheses on the last digits, are one standard
deviation.}
\label{tab:fig3}
\centering\footnotesize
\setlength{\tabcolsep}{3.4pt}
\begin{tabular}{l c c c c c c c c c c}
\toprule
\multicolumn{11}{l}{\textit{Frequencies}: $\nu=[a+b\coth(\Theta/2T)]^{1/2}+\lambda\Phi$}\\[2pt]
 & $\Phi$ & Range (K) & $N$ & $a$ (THz$^2$) & $b$ (THz$^2$) & $\Theta$ (K) &
 $\lambda$ (GHz) & $\Delta\mathrm{BIC}_\mathrm{SL}$ & $\Delta\mathrm{BIC}_\Phi$ & rms (GHz)\\
\midrule
$\nu_1$  & LRO & 1.6--301 & 31 & 0.6127(80) & $-0.0827(99)$ & 310(27) & $-9.8(14)$  & 27.5   & 21.3 & 1.1\\
$\nu_1'$ & LRO & 20--301  & 29 & 0.6076(45) & $-0.0758(68)$ & 184(14) & $+28.8(16)$ & 80.0   & 41.7 & 1.6\\
$\nu_2$  & --  & 1.6--191 & 20 & 4.370(23)  & $-0.007(64)$  & 5\tn{a} & --          & $-1.7$ & --   & 9.4\\
$\nu_2'$ & LRO & 20--301  & 29 & 6.741(90)  & $-1.67(13)$   & 208(14) & $+102(9)$   & 52.8   & 22.9 & 9.1\\
\midrule
\multicolumn{11}{l}{\textit{Linewidths}: $\gamma=\gamma_0+c\coth(\Theta_\gamma/2T)+\eta\Phi$}\\[2pt]
 & $\Phi$ & Range (K) & $N$ & $\gamma_0$ (THz) & $c$ (THz) & $\Theta_\gamma$ (K) &
 $\eta$ (THz) & $\Delta\mathrm{BIC}_\mathrm{SL}$ & $\Delta\mathrm{BIC}_\Phi$ & rms (GHz)\\
\midrule
$\gamma_1$  & SRO & 1.6--301 & 31 & 0.11(28)\tn{b} & 0.05(35)\tn{b} & 500\tn{a} & $+0.205(71)$ & 95.4 & 29.9 & 11\\
$\gamma_1'$ & SRO & 20--301  & 29 & 0.254(12)      & 0.001(11)      & --\tn{c}  & $-0.101(10)$ & 44.1 & 21.2 & 4.0\\
$\gamma_2$  & --  & 1.6--191 & 20 & 0\tn{a}        & 0.255(88)      & 254(46)   & --           & 0.8  & --   & 12\\
$\gamma_2'$ & step\tn{d} & 20--231 & 22 & 0\tn{a}  & 0.070(85)      & 152(190)  & $+0.259(95)$ & 9304 & --   & 7.7\\
\midrule
\multicolumn{11}{l}{\textit{Oscillator strengths}: $A=A_0+c_AT+\zeta\Phi$ \quad ($A_0$, $\zeta$ and rms in $\Omega^{-1}\,\mathrm{cm}^{-1}\,\mathrm{THz}$; $c_A$ in $10^{-4}~\Omega^{-1}\,\mathrm{cm}^{-1}\,\mathrm{THz}\,\mathrm{K}^{-1}$)}\\[2pt]
 & $\Phi$ & Range (K) & $N$ & $A_0$ & $c_A$ & & $\zeta$ & $\Delta\mathrm{BIC}_\mathrm{SL}$ &
 $\Delta\mathrm{BIC}_\Phi$ & rms ($10^{-3}$)\\
\midrule
$A_1$  & SRO & 1.6--301 & 31 & 0.133(17)  & $-0.30(51)$ & & $+0.052(15)$   & 7.2  & 10.7 & 6.8\\
$A_1'$ & SRO & 20--301  & 29 & 0.166(12)  & $-1.13(36)$ & & $+0.061(10)$   & 21.1 & 4.3  & 4.3\\
$A_2$  & SRO & 1.6--231 & 24 & 0.060(31)  & $-1.4(10)$  & & $+0.146(26)$   & 18.5 & 20.7 & 9.9\\
$A_2'$ & LRO\tn{e} & 20--231 & 22 & 0.1839(62) & $+1.62(37)$ & & $-0.0323(44)$ & 712 & 564 & 5.7\\
\bottomrule
\end{tabular}

\smallskip
\begin{minipage}{\linewidth}\footnotesize
\tn{a}\,At the bound of the search interval ($\Theta$ and $\Theta_\gamma$ between 5
and 500~K; $\gamma_0\geq0$). For $\nu_2$, $\Theta=5$~K reduces the background to
$\nu^2\approx a+2bT/\Theta$, linear in $T$, of which only $b/\Theta$ is
determined.\quad
\tn{b}\,Strongly correlated with each other and with $\Theta_\gamma$.\quad
\tn{c}\,$c$ is compatible with zero, and $\Theta_\gamma$ is undetermined
($41\pm743$~K).\quad
\tn{d}\,Background plus $\eta S(T)$, with $S(T)=[1+e^{(T-T_c)/w}]^{-1}$,
$T_c=110(5)$~K and $w=21(6)$~K; weighted fit, $\chi^2_\nu=15.1$; rms over the
points without systematic uncertainty.\quad
\tn{e}\,Weighted fit, $\chi^2_\nu=13.6$; rms over the points without systematic
uncertainty.
\end{minipage}
\end{table}

\begin{table}[htbp]
\caption{\textbf{Robustness of the temperature models to the lattice of the Ising
functions.} BIC decrease produced by the spin--lattice term and its amplitude
($\lambda$ in GHz, $\eta$ in THz, $\zeta$ in
$\Omega^{-1}\,\mathrm{cm}^{-1}\,\mathrm{THz}$) with the functions of the honeycomb
lattice, used in the main text, and of the square lattice. The selected form of
$\Phi$ is the same with both lattices. For $\nu_2$ and $\gamma_2$, no term is
retained with either lattice.}
\label{tab:lattice}
\centering\small
\begin{tabular}{l c c c c c}
\toprule
 & $\Phi$ & \multicolumn{2}{c}{$\Delta\mathrm{BIC}_\mathrm{SL}$} & \multicolumn{2}{c}{Amplitude}\\
\cmidrule(lr){3-4}\cmidrule(lr){5-6}
 & & Honeycomb & Square & Honeycomb & Square\\
\midrule
$\nu_1$     & LRO & 27.5 & 27.3 & $\lambda=-9.8(14)$   & $-10.2(14)$\\
$\nu_1'$    & LRO & 80.0 & 80.2 & $\lambda=+28.8(16)$  & $+29.7(17)$\\
$\nu_2'$    & LRO & 52.8 & 52.6 & $\lambda=+102(9)$    & $+105(10)$\\
$\gamma_1$  & SRO & 95.4 & 99.8 & $\eta=+0.205(71)$    & $+0.172(54)$\\
$\gamma_1'$ & SRO & 44.1 & 41.8 & $\eta=-0.101(10)$    & $-0.083(11)$\\
$A_1$       & SRO & 7.2  & 2.7  & $\zeta=+0.052(15)$   & $+0.034(14)$\\
$A_1'$      & SRO & 21.1 & 19.0 & $\zeta=+0.061(10)$   & $+0.0481(87)$\\
$A_2$       & SRO & 18.5 & 12.7 & $\zeta=+0.146(26)$   & $+0.116(26)$\\
$A_2'$      & LRO & 712  & 726  & $\zeta=-0.0323(44)$  & $-0.0331(44)$\\
$\nu_2$     & --  & $-1.7$ & $-2.3$ & -- & --\\
$\gamma_2$  & --  & 0.8  & $-0.1$ & -- & --\\
\bottomrule
\end{tabular}
\end{table}

\snote{Temperature scale of the spin correlations in the oscillator strengths}\label{sn:scale}

With $T_\mathrm{N}$ fixed at 118~K, the $\Phi_\mathrm{SRO}$ terms of the
oscillator strengths at $\theta=0^\circ$ leave systematic residuals (Fig.~3c,f of
the main text). $A_1$ increases by 7\% between 1.6 and 91~K, whereas the model
decreases monotonically: it lies above the data up to 20~K, with residuals of
$-9$ to $-11\times10^{-3}~\Omega^{-1}\,\mathrm{cm}^{-1}\,\mathrm{THz}$, and below
them between 121 and 151~K, with residuals of $+10$ to $+14\times10^{-3}$. Above
200~K, the description is adequate, with
a root-mean-square residual of $4.6\times10^{-3}$. The model of $A_2$ likewise
decreases faster than the data just above $T_\mathrm{N}$ (residuals of $+10$ to
$+22\times10^{-3}$ between 121 and 161~K) and lies above them between 210 and
231~K (residuals of $-11$ to $-19\times10^{-3}$). In both cases, the measured
decrease is centred above $T_\mathrm{N}$.

To separate the form of the correlation function from its temperature scale, we
replaced $T_\mathrm{N}$ in $\Phi_\mathrm{SRO}$ by a free scale $T^*$,
$\Phi_\mathrm{SRO}(T;T^*)=\Phi_\mathrm{SRO}(T\,T_\mathrm{N}/T^*)$, which is
equivalent to $K=K_\mathrm{c}T^*/T$, and varied $T^*$ from 90 to 230~K in steps
of 0.5~K, with the same points as in the fits of Fig.~3. The intervals given below
contain the values of $T^*$ for which $\chi^2$, scaled by the residual variance,
lies within 1 of its minimum.
\begin{itemize}
\item $A_1$: $T^*=152.5$~K (150.5--155.0~K). The BIC decreases by 59 with respect
to the fit with $T_\mathrm{N}$, and the root-mean-square residual falls to
(2.2--2.8)$\times10^{-3}$ in all temperature ranges.
\item $A_2$: $T^*=142.5$~K (140.5--145.0~K), with a BIC decrease of 38.
\item $A_1'$: $T^*=118.5$~K (114.0--121.5~K), with no improvement over the fit with
$T_\mathrm{N}$ (BIC higher by 3.3).
\end{itemize}
With the square-lattice functions, $T^*=158$, 146 and 120~K, respectively.

The form of $\Phi_\mathrm{SRO}$ is thus adequate for $A_1$ and $A_2$, but their
temperature scale lies 20--30\% above $T_\mathrm{N}$, whereas $A_1'$ follows
$T_\mathrm{N}$. The strengths at $\theta=0^\circ$ therefore respond to spin
correlations that develop well above the ordering temperature, on a scale that
the nearest-neighbour Ising model, with $T_\mathrm{N}$ as its only scale, does not
produce. The bond-dependent correlations of the $J_1$--$J_2$--$J_3$ Hamiltonian of
the zigzag order~\cite{Lancon2016}, which no nearest-neighbour model represents,
are one ingredient that a description of this scale would have to include. The
two strengths at $\theta=0^\circ$ are, moreover, closely linked: between 91 and
191~K, $A_1=0.115+0.387A_2$, with a correlation coefficient of 0.996, so that the
decrease of $A_1$ above $T_\mathrm{N}$ accompanies that of $A_2$. The fits with a
free scale serve only as a diagnostic; all curves in Fig.~3 use $T_\mathrm{N}$.

\snote{Critical-field model of mode~2 at $\theta=0^\circ$}\label{sn:field}

This note describes the model used for $\nu_2$, $\gamma_2$ and $A_2$ at
$\theta=0^\circ$ and 1.6~K (Fig.~4c,d,g of the main text). The parameters of both
modes obtained from the field series are listed in Supplementary
Table~\ref{tab:field}, where $\gamma_2$ is the FWHM of the fitted line and $A_2$
its total strength (Supplementary Note~\ref{sn:fits}). Mode~1 shows no anomaly at
$B^*$. Between 0 and 7~T, $\nu_1$ increases by 5~GHz and $\gamma_1$ decreases by
0.065~THz, both smoothly, and $A_1$ varies by less than 3\%; a change of slope of
$A_1$ at $B^*$ does not improve its fit (BIC of 11.5, against 10.8 for a straight
line).

\begin{table}[p]
\caption{\textbf{Oscillator parameters of the field series at $\theta=0^\circ$
and 1.6~K} ($\mathbf{B}\parallel c^*$). $\nu_i$: resonance frequency; $\gamma_i$:
linewidth (FWHM); $A_i$: oscillator strength, in
$\Omega^{-1}\,\mathrm{cm}^{-1}\,\mathrm{THz}$. For mode~2, $\nu_2$ is the common
centre of the components, $\gamma_2$ the FWHM of the summed line and $A_2$ its
total strength; $n_2$ is the number of components (Supplementary
Note~\ref{sn:fits}). Uncertainties, in parentheses on the last digits, are one
standard deviation.}
\label{tab:field}
\centering\small
\begin{tabular}{c c c c c c c c}
\toprule
$B$ (T) & $\nu_1$ (THz) & $\gamma_1$ (THz) & $A_1$ & $\nu_2$ (THz) & $\gamma_2$ (THz) & $A_2$ & $n_2$\\
\midrule
0.00 & 0.7064(19) & 0.3351(64) & 0.1986(31) & 1.97757(88) & 0.1666(48) & 0.2070(47) & 2 \\
0.25 & 0.7064(18) & 0.3343(64) & 0.1984(31) & 1.97570(85) & 0.1646(44) & 0.2095(47) & 2 \\
0.50 & 0.7065(18) & 0.3330(64) & 0.1986(31) & 1.97465(85) & 0.1604(44) & 0.2111(48) & 2 \\
0.75 & 0.7066(18) & 0.3324(63) & 0.1989(31) & 1.97245(85) & 0.1576(44) & 0.2099(47) & 2 \\
1.00 & 0.7068(18) & 0.3305(62) & 0.1988(31) & 1.97176(83) & 0.1508(42) & 0.2147(50) & 2 \\
1.25 & 0.7071(18) & 0.3286(62) & 0.1987(31) & 1.96982(79) & 0.1418(40) & 0.2167(49) & 2 \\
1.50 & 0.7072(18) & 0.3271(61) & 0.1988(30) & 1.96774(77) & 0.1376(39) & 0.2174(50) & 2 \\
1.75 & 0.7072(18) & 0.3256(61) & 0.1993(31) & 1.96542(76) & 0.1360(39) & 0.2191(49) & 2 \\
2.00 & 0.7076(18) & 0.3241(61) & 0.1997(30) & 1.96374(72) & 0.1252(38) & 0.2175(48) & 2 \\
2.25 & 0.7076(18) & 0.3224(60) & 0.1996(30) & 1.96304(69) & 0.1186(37) & 0.2219(50) & 2 \\
2.50 & 0.7080(17) & 0.3207(59) & 0.1996(30) & 1.95985(64) & 0.1070(35) & 0.2235(48) & 2 \\
2.75 & 0.7082(17) & 0.3186(58) & 0.1994(30) & 1.95844(59) & 0.0944(35) & 0.2233(48) & 2 \\
3.00 & 0.7085(17) & 0.3170(58) & 0.2002(30) & 1.95582(49) & 0.0722(32) & 0.2253(48) & 2 \\
3.25 & 0.7086(17) & 0.3147(57) & 0.2003(30) & 1.95387(39) & 0.0512(30) & 0.2232(47) & 2 \\
3.50 & 0.7089(17) & 0.3116(57) & 0.1993(30) & 1.94749(59) & 0.0866(36) & 0.2241(51) & 2 \\
3.75 & 0.7090(16) & 0.3092(56) & 0.1994(30) & 1.94267(57) & 0.1022(31) & 0.2219(44) & 2 \\
4.00 & 0.7085(16) & 0.3090(56) & 0.2004(30) & 1.93794(60) & 0.1132(35) & 0.2171(40) & 2 \\
4.25 & 0.7089(16) & 0.3062(56) & 0.2001(30) & 1.93197(62) & 0.1340(44) & 0.2136(34) & 2 \\
4.50 & 0.7091(16) & 0.3033(55) & 0.2000(30) & 1.92838(64) & 0.1440(57) & 0.2129(32) & 2 \\
4.75 & 0.7093(16) & 0.3009(54) & 0.2000(30) & 1.92176(67) & 0.1586(55) & 0.2076(28) & 2 \\
5.00 & 0.7097(16) & 0.2978(54) & 0.1994(30) & 1.91683(71) & 0.1774(23) & 0.2054(22) & 1 \\
5.25 & 0.7100(16) & 0.2952(53) & 0.1995(30) & 1.91135(68) & 0.1798(22) & 0.2031(21) & 1 \\
5.50 & 0.7101(15) & 0.2923(53) & 0.1995(30) & 1.90478(67) & 0.1864(21) & 0.2010(20) & 1 \\
5.75 & 0.7103(15) & 0.2887(52) & 0.1988(29) & 1.89949(63) & 0.1912(20) & 0.1995(18) & 1 \\
6.00 & 0.7105(15) & 0.2848(51) & 0.1982(29) & 1.89390(63) & 0.1964(20) & 0.1977(17) & 1 \\
6.25 & 0.7107(15) & 0.2814(51) & 0.1978(29) & 1.88875(61) & 0.1996(19) & 0.1963(16) & 1 \\
6.50 & 0.7110(15) & 0.2768(50) & 0.1965(29) & 1.88365(60) & 0.2030(19) & 0.1961(16) & 1 \\
6.75 & 0.7111(15) & 0.2739(50) & 0.1964(30) & 1.87917(58) & 0.2072(18) & 0.1933(15) & 1 \\
7.00 & 0.7114(15) & 0.2703(49) & 0.1951(29) & 1.87546(58) & 0.2104(18) & 0.1923(15) & 1 \\
\bottomrule
\end{tabular}
\end{table}

\paragraph{Statics.}
We describe the change at $B^*$ by a Landau
expansion~\cite{LandauLifshitz1980,ChaikinLubensky1995} in an order parameter
$\psi$, whose microscopic identity is left open, coupled to the normal coordinate
$Q$ of mode~2, with reduced mass $\mu$ and bare frequency $\omega_0$:
\begin{equation}
F=F_0+\tfrac12r(B)\psi^2+\tfrac14u\psi^4+\tfrac12\mu\omega_0^2Q^2+\tfrac12\kappa\psi^2Q^2,
\label{eq:landau}
\end{equation}
with $u>0$. The field enters $r$ through the dependence on $\psi$ of the
magnetization, $M=m_0+m_2\psi^2$, and of the susceptibility,
$\chi=\chi_0+\chi_2\psi^2$: the Zeeman terms $-MB-\frac12\chi B^2$ give
$r(B)=r_0-2m_2B-\chi_2B^2$. The linear route requires a net moment that changes
with $\psi$, whereas the quadratic route exists also for layers without a net
moment along $c^*$, such as the zigzag layers of FePS$_3$. Near the field $B^*$
at which $r$ vanishes, $r\approx r_1(B^*-B)$ with $r_1>0$, and minimization of
$F$ gives $\psi=0$ below $B^*$ and $\langle\psi^2\rangle=(r_1/u)(B-B^*)$ above.
The term $\frac12\kappa\psi^2Q^2$, the lowest-order coupling allowed for a mode
that is not totally symmetric, renormalizes the force constant,
$\mu\omega_2^2=\mu\omega_0^2+\kappa\langle\psi^2\rangle$, so that
\begin{equation}
\nu_2(B)=\nu_0-s_0B-s_1(B-B^*)_+,\qquad s_1=-\frac{\kappa r_1}{8\pi^2\mu\nu_2u},
\label{eq:nu2B}
\end{equation}
where $(x)_+=\max(x,0)$. $\nu_2$ is continuous at $B^*$ and changes slope there,
and $s_1>0$ corresponds to $\kappa<0$. The linear term $-s_0B$, present already at
low field, does not arise from $\psi$, and its origin is not determined by our
data. Since the departure of $\nu_2$ from its low-field line is proportional to
$\langle\psi^2\rangle$, it grows as $(B-B^*)_+^p$ with $p=2\beta$, where $\beta$
is the order-parameter exponent; Landau theory gives $p=1$.

If the order parameter also changes the dipole of the mode, its effective charge
becomes $Z=Z^*+z_1\psi^2+z_2\psi^4$. Since the oscillator strength is proportional
to $Z^2/\mu$~\cite{GonzeLee1997,Baroni2001},
\begin{equation}
A_2(B)=A_0+k_0B+k_1(B-B^*)_++k_2(B-B^*)_+^2,
\label{eq:A2B}
\end{equation}
with $k_1=2A^*(z_1/Z^*)(r_1/u)$ and $k_2=k_2^0+2A^*(z_2/Z^*)(r_1/u)^2$, where
$A^*=A_2(B^*)$ and $k_2^0=k_1^2/(4A^*)$ is the part that arises from the square of
$Z$ alone. As for $\nu_2$, the linear term $k_0B$ does not arise from $\psi$. The
ratio of the relative changes of slope of $A_2$ and $\nu_2$ at $B^*$,
\begin{equation}
R=\frac{k_1/A_2(B^*)}{-s_1/\nu_2(B^*)}
=\frac{\partial\ln A_2/\partial\langle\psi^2\rangle}{\partial\ln\nu_2/\partial\langle\psi^2\rangle},
\label{eq:R}
\end{equation}
does not depend on the unknown amplitude $r_1/u$. Because $\nu_2\propto(K/\mu)^{1/2}$
and $A_2\propto Z^2/\mu$, with $K$ the force constant of the mode,
$R/4=(\partial\ln Z/\partial\langle\psi^2\rangle)/(\partial\ln K/\partial\langle\psi^2\rangle)$
compares the sensitivities of the effective charge and of the force constant to the
order parameter.

\paragraph{Dynamics.}
The linewidth is set by the relaxation of $\psi$. For relaxational
(Landau--Khalatnikov) dynamics~\cite{Rehwald1973,HohenbergHalperin1977},
$\partial\psi/\partial t=-\Gamma_0\,\partial F/\partial\psi$, small deviations
from equilibrium decay at the rate $\Gamma_\psi=\Gamma_0\,\partial^2F/\partial\psi^2$,
which equals $\Gamma_0r_1(B^*-B)$ below $B^*$ and $2\Gamma_0r_1(B-B^*)$ above it:
the relaxation time $\tau=1/\Gamma_\psi$ diverges at $B^*$ (critical slowing down).
A coupling $g\,\delta\psi\,Q$, linear in $Q$, adds to the equation of motion of the
mode the self-energy $\Pi(\omega)$:
\begin{equation}
\omega^2=\omega_0^2-i\omega\gamma_0+\Pi(\omega),\qquad
\Pi(\omega)=-\frac{\Delta^2}{1-i\omega\tau},\qquad \Delta^2=\frac{g^2\chi_\psi}{\mu},
\label{eq:selfenergy}
\end{equation}
with $\chi_\psi=(\partial^2F/\partial\psi^2)^{-1}$. The resulting contributions to
the linewidth (FWHM, in angular frequency) and to the frequency of the mode are
\begin{equation}
\delta\gamma=\frac{\Delta^2\tau}{1+\omega_2^2\tau^2},\qquad
\delta\omega=-\frac{\Delta^2}{2\omega_2(1+\omega_2^2\tau^2)}.
\label{eq:dgamma}
\end{equation}
In ultrasonic experiments, $\omega\tau\ll1$, and $\delta\gamma\approx\Delta^2\tau$
has a maximum at the transition: the anomalous attenuation of sound near
structural phase transitions~\cite{Rehwald1973}. For a terahertz mode,
$\omega_2\tau\gg1$, and
\begin{equation}
\delta\gamma\approx\frac{\Delta^2\Gamma_\psi}{\omega_2^2}\propto|B-B^*|:
\label{eq:cusp}
\end{equation}
the damping contributed by the order parameter vanishes at $B^*$, where the order
parameter relaxes most slowly, and the linewidth has a cusp-like minimum. The
corresponding frequency shift, $\delta\omega\approx-\Delta^2/(2\omega_2^3\tau^2)$, is
of second order in $B-B^*$, so that near $B^*$ the frequency records only the
static change of slope of Eq.~(\ref{eq:nu2B}). The same dynamics changes the area
of the line only at order $(\delta\gamma/\omega_2)/(\omega_2\tau)$, by less than
1\% for $\omega_2\tau\gtrsim8$; the anomaly of $A_2$ is therefore static. The
linewidth was fitted with
\begin{equation}
\gamma_2(B)=\gamma^*+\frac{\Gamma_{\lessgtr}\,x}{1+x/x_{\lessgtr}},\qquad x=|B-B^*|,
\label{eq:gamma2B}
\end{equation}
with separate slopes, $\Gamma_<$ below and $\Gamma_>$ above $B^*$, and saturation
fields $x_<$ and $x_>$ that describe, without a microscopic model, the flattening
of the flanks away from $B^*$.

Three remarks qualify this description. First, where
$\omega_2\tau\approx1$, Eq.~(\ref{eq:dgamma}) gives $\delta\omega=-\delta\gamma/2$:
if the flattening of the flanks came from $\omega_2\tau\approx1$, it would be
accompanied by a softening of 60--80~GHz, half the magnetic contribution to the
linewidth, which is not observed. Hence $\omega_2\tau\gg1$ over the whole field
range, that is, $\tau\gg1/\omega_2\approx0.08$~ps, and the flattening must come
from a field dependence of the coupling. Second, relaxational damping requires a
coupling linear in $Q$. Above $B^*$, a term $\kappa'\psi^2Q$ gives
$g=2\kappa'\langle\psi\rangle$ and, in mean field, a constant
$\Delta^2=2\kappa'^2/(u\mu)$; such a term is allowed only if $Q$ is totally
symmetric. Below $B^*$, $\langle\psi\rangle=0$, and the linear increase of
$\gamma_2$ requires another channel, such as fluctuations of $\psi$, which the
minimal form does not fix. Third, with the same $\Delta^2$ on both sides,
mean-field theory gives a flank twice as steep above $B^*$ as below, whereas the
measured ratio is $\Gamma_>/\Gamma_<=1.1\pm0.2$: the coupling differs on the two
sides. The identity and symmetry of $\psi$, which determine the allowed couplings,
remain open.

\paragraph{Fits.}
The models were fitted to the values of Supplementary Table~\ref{tab:field} by
weighted least squares, with the uncertainties of the oscillator fits. The
uncertainties of the parameters were scaled by $\max(\chi^2_\nu,1)^{1/2}$, and the
95\% bands of Fig.~4 were obtained as in Supplementary Note~\ref{sn:selection};
for $A_2$, the uncertainty of $B^*$ was added in quadrature. Models fitted to the
same data were compared through $\mathrm{BIC}=\chi^2+k\ln N$, with $k$ the number
of parameters. Because the uncertainties of $A_2$ (median 0.0047) are about four
times the scatter between neighbouring points (0.0011), the fits of $A_2$ have
$\chi^2_\nu<1$; we therefore also tested the randomness of their residuals through
the number of runs of equal sign~\cite{WaldWolfowitz1940}, which does not depend on
the scale of the uncertainties.

\paragraph{Results.}
The joint fit of Eqs.~(\ref{eq:nu2B}) and (\ref{eq:gamma2B}) to $\nu_2$ and
$\gamma_2$, with a common critical field, gives $B^*=3.217\pm0.024$~T and describes
both quantities within their uncertainties ($\chi^2_\nu=1.44$; root-mean-square
residuals of 0.74~GHz in $\nu_2$ and 3.6~GHz in $\gamma_2$; Supplementary
Table~\ref{tab:Bparams}). At $B^*$, the slope of $\nu_2$ changes from $-7.45$ to
$-21.39$~GHz~T$^{-1}$, and $\gamma_2$ reaches its residual width,
$\gamma^*=0.047\pm0.004$~THz, with flanks of 132 and 147~GHz~T$^{-1}$. Separate
fits place the change of slope of $\nu_2$ at $3.18\pm0.03$~T and the minimum of
$\gamma_2$ at $3.25\pm0.03$~T. Above $B^*$, the departure of $\nu_2$ from its
low-field line grows with $p=0.972\pm0.021$ ($0.979\pm0.027$ with $B^*$ free, at
$3.197\pm0.043$~T), the mean-field value; the exponents of the two- and
three-dimensional Ising models~\cite{Onsager1944,Yang1952,PelissettoVicari2002} and
of a mean-field tricritical point~\cite{ChaikinLubensky1995} are excluded
(Supplementary Table~\ref{tab:Bcomp}).

For $A_2$, with $B^*$ fixed, the first-order form of Eq.~(\ref{eq:A2B}) ($k_2=0$)
describes the trend but leaves a pattern in the residuals, positive between 1 and
4~T and negative between 4.75 and 6~T (runs test, $z=-2.8$). The second-order
term removes the pattern ($z=-1.7$) and lowers the BIC by 6.2. A saturating
decrease above $B^*$ is equivalent to it, whereas smooth forms without $B^*$, and a
dependence on $\langle\psi\rangle$ instead of $\langle\psi^2\rangle$, that is, on
$(B-B^*)_+^{1/2}$, are excluded (Supplementary Table~\ref{tab:Bcomp}). Fitted alone
with the second-order form, $A_2$ changes slope at 3.34~T (3.02--3.52~T), and a
joint fit of $\nu_2$, $\gamma_2$ and $A_2$ gives $B^*=3.217\pm0.020$~T
($\chi^2_\nu=0.99$); a separate critical field for $A_2$ is disfavoured (BIC
higher by 4.3). The relative changes of slope at $B^*$,
$k_1/A_2(B^*)=-8.9\pm1.2\%$~T$^{-1}$ and $-s_1/\nu_2(B^*)=-0.713\pm0.014\%$~T$^{-1}$,
give $R=12.5\pm1.7$ ($7.9\pm0.8$ with the first-order form): the effective charge
of mode~2 responds to the order parameter two to three times more strongly than its
force constant ($R/4=3.1$ and 2.0). The second-order coefficient,
$k_2=(1.36\pm0.44)\times10^{-3}$, exceeds the part that arises from the square of
the effective charge alone, $k_2^0=(0.45\pm0.12)\times10^{-3}$ (both in
$\Omega^{-1}\,\mathrm{cm}^{-1}\,\mathrm{THz}\,\mathrm{T}^{-2}$), by 2.7 standard
deviations, which points to a term in $\psi^4$ in the effective charge.

\paragraph{Alternative descriptions.}
Supplementary Table~\ref{tab:Bcomp} also compares the critical-field model with
three alternatives, fitted to the same values of $\nu_2$ and $\gamma_2$.
(i)~A smooth crossover between two field-dependent relaxation channels, in which a
canting angle $\varphi_\mathrm{c}$, with $\sin\varphi_\mathrm{c}=B/B_E$,
renormalizes the frequency, $\nu_2=\nu_0(1+2\lambda_B\sin^2\varphi_\mathrm{c})^{1/2}$,
and the linewidth is
$\gamma_2=\gamma_\mathrm{res}+\gamma_\mathrm{I}/[1+(B/B_\mathrm{I})^2]+\gamma_\mathrm{II}\sin^2\varphi_\mathrm{c}$:
$\chi^2_\nu=35.9$, with root-mean-square residuals of 2.7~GHz in $\nu_2$ and 19~GHz
in $\gamma_2$. A smooth form produces neither the change of slope of $\nu_2$ nor
the cusp of $\gamma_2$. (ii)~A resonance with an excitation whose frequency crosses
that of the mode at $B^*$, with a perturbative self-energy
$\Sigma=V^2/(\delta-i\Gamma_X)$, where $V$ is the coupling, $\Gamma_X$ the width of
the partner and $\delta=s_X(B-B^*)$ the detuning, and quadratic backgrounds in
$\nu_2$ and $\gamma_2$: $\chi^2_\nu=31$--34 for the two directions of crossing.
When the partner is narrower than the mode, $\mathrm{Im}\,\Sigma$ narrows the mode
at the crossing, but $\mathrm{Re}\,\Sigma$ then shifts its centre dispersively, by
one quarter of the depth of the dip, that is, by $\approx\pm25$--30~GHz for the
measured dip; no such excursion is observed. (iii)~Two coupled modes, described by
the exact complex eigenvalue of the phonon-like branch of a non-Hermitian
$2\times2$ matrix~\cite{BarkerHopfield1964}: $\chi^2_\nu=21.6$ (best of 108
starting points). At the crossing, the widths of the two branches tend to the mean
of the bare widths, and the phonon-like branch does not narrow below about half of
its background width: its minimum width is 0.10~THz, against the measured
0.047~THz. Neither a smooth crossover nor a resonant crossing reproduces the narrow
cusp of $\gamma_2$.

\begin{table}[htbp]
\caption{\textbf{Parameters of the critical-field model} ($\theta=0^\circ$,
1.6~K; Supplementary Note~\ref{sn:field}). $A_2$ and its coefficients in
$\Omega^{-1}\,\mathrm{cm}^{-1}\,\mathrm{THz}$, per T or per T$^2$. Uncertainties
are one standard deviation.}
\label{tab:Bparams}
\centering\small
\begin{tabular}{l l c c}
\toprule
\multicolumn{4}{l}{\textit{(a) Joint fit of $\nu_2$ and $\gamma_2$, Eqs.~(\ref{eq:nu2B}) and
(\ref{eq:gamma2B}); 58 points, 9 parameters, $\chi^2_\nu=1.44$}}\\
\midrule
$B^*$ & critical field (T) & \multicolumn{2}{c}{$3.217\pm0.024$}\\
$\nu_0$ & $\nu_2$ extrapolated to $B=0$ (THz) & \multicolumn{2}{c}{$1.97847\pm0.00049$}\\
$s_0$ & $-\mathrm{d}\nu_2/\mathrm{d}B$ below $B^*$ (GHz~T$^{-1}$) & \multicolumn{2}{c}{$7.45\pm0.21$}\\
$s_1$ & change of $-\mathrm{d}\nu_2/\mathrm{d}B$ at $B^*$ (GHz~T$^{-1}$) & \multicolumn{2}{c}{$13.94\pm0.28$}\\
$s_0+s_1$ & $-\mathrm{d}\nu_2/\mathrm{d}B$ above $B^*$ (GHz~T$^{-1}$) & \multicolumn{2}{c}{$21.39\pm0.13$}\\
$\gamma^*$ & $\gamma_2$ at $B^*$ (THz) & \multicolumn{2}{c}{$0.0466\pm0.0040$}\\
$\Gamma_<$; $x_<$ & flank below $B^*$ (GHz~T$^{-1}$; T) & \multicolumn{2}{c}{$132\pm23$; $1.23\pm0.27$}\\
$\Gamma_>$; $x_>$ & flank above $B^*$ (GHz~T$^{-1}$; T) & \multicolumn{2}{c}{$147\pm11$; $1.59\pm0.15$}\\
\midrule
\multicolumn{4}{l}{\textit{(b) $A_2$, Eq.~(\ref{eq:A2B}) with $B^*$ fixed at the value of (a); 29 points}}\\
\midrule
 & & First order ($k_2=0$) & Second order\\
\cmidrule(lr){3-3}\cmidrule(lr){4-4}
$A_0$ & & $0.2096\pm0.0023$ & $0.2077\pm0.0023$\\
$k_0$ & ($10^{-3}$~T$^{-1}$) & $4.04\pm0.92$ & $5.95\pm1.11$\\
$k_1$ & ($10^{-3}$~T$^{-1}$) & $-12.5\pm1.3$ & $-20.2\pm2.8$\\
$k_2$ & ($10^{-3}$~T$^{-2}$) & -- & $1.36\pm0.44$\\
$\chi^2_\nu$ & & 0.48 & 0.11\\
$A_2(B^*)$ & & $0.2226\pm0.0014$ & $0.2269\pm0.0020$\\
$k_1/A_2(B^*)$ & (\%~T$^{-1}$) & $-5.6\pm0.6$ & $-8.9\pm1.2$\\
$R$ & Eq.~(\ref{eq:R})\tn{a} & $7.9\pm0.8$ & $12.5\pm1.7$\\
\midrule
\multicolumn{4}{l}{\textit{(c) Joint fit of $\nu_2$, $\gamma_2$ and $A_2$ with a common $B^*$; 87 points}}\\
\midrule
$B^*$ & (T) & $3.214\pm0.021$ & $3.217\pm0.020$\\
$\chi^2_\nu$ & & 1.11 & 0.99\\
$\Delta\mathrm{BIC}$ & separate $B^*$ for $A_2$\tn{b} & $+1.4$ & $+4.3$\\
\bottomrule
\end{tabular}

\smallskip
\begin{minipage}{0.9\linewidth}\footnotesize
\tn{a}\,With $-s_1/\nu_2(B^*)=-0.713\pm0.014\%$~T$^{-1}$ and $\nu_2(B^*)=1.9545$~THz.\quad
\tn{b}\,BIC increase when $A_2$ is given its own critical field.
\end{minipage}
\end{table}

\begin{table}[htbp]
\caption{\textbf{Comparison of models for mode~2 in field} ($\theta=0^\circ$,
1.6~K; Supplementary Note~\ref{sn:field}). $k$: number of parameters. Runs:
number of runs of equal sign in the residuals, with the value expected for random
residuals and the corresponding $z$ score~\cite{WaldWolfowitz1940}.}
\label{tab:Bcomp}
\centering\small
\begin{tabular}{l c c}
\toprule
\multicolumn{3}{l}{\textit{(a) $\nu_2$ and $\gamma_2$, 58 points}}\\
Model & $k$ & $\chi^2_\nu$\\
\midrule
Critical field, Eqs.~(\ref{eq:nu2B}) and (\ref{eq:gamma2B}) & 9 & 1.44\\
Smooth crossover of two relaxation channels & 7 & 35.9\\
Resonance with a crossing excitation, perturbative\tn{a} & 9 & 31--34\\
Two coupled modes, exact eigenvalue & 10 & 21.6\\
\bottomrule
\end{tabular}

\medskip
\begin{tabular}{c l c}
\toprule
\multicolumn{3}{l}{\textit{(b) Exponent $p$ of the departure of $\nu_2$ above $B^*$; 29 points, 3 parameters, $B^*$ fixed}}\\
$p$ & Reference value & $\chi^2$\\
\midrule
1/4 & two-dimensional Ising, $2\beta=1/4$ & 3577\\
1/2 & mean-field tricritical point & 1425\\
0.652 & three-dimensional Ising & 557\\
1 & mean field & 38.0\\
1.5 & -- & 457\\
$0.972\pm0.021$ & fitted (4 parameters) & 35.7\\
\bottomrule
\end{tabular}

\medskip
\begin{tabular}{l c c c c}
\toprule
\multicolumn{5}{l}{\textit{(c) $A_2$ with $B^*$ fixed, 29 points}}\\
Form & $k$ & $\chi^2$ & BIC & Runs (expected; $z$)\\
\midrule
Straight line & 2 & 104.8 & 111.5 & 3 (15.5; $-4.7$)\\
Quadratic in $B$ & 3 & 35.3 & 45.4 & 4 (15.1; $-4.3$)\\
Cubic in $B$ & 4 & 11.3 & 24.8 & 7 (15.5; $-3.2$)\\
Eq.~(\ref{eq:A2B}), first order & 3 & 12.4 & 22.5 & 8 (15.3; $-2.8$)\\
Eq.~(\ref{eq:A2B}), second order & 4 & 2.8 & 16.3 & 11 (15.5; $-1.7$)\\
Saturating decrease above $B^*$\tn{b} & 4 & 3.0 & 16.4 & 11 (15.5; $-1.7$)\\
First order plus a quadratic background in $B$ & 4 & 4.6 & 18.1 & 11 (14.7; $-1.5$)\\
$(B-B^*)_+^{1/2}$ instead of $(B-B^*)_+$ & 3 & 46.4 & 56.5 & 3 (14.1; $-4.7$)\\
\bottomrule
\end{tabular}

\smallskip
\begin{minipage}{0.9\linewidth}\footnotesize
\tn{a}\,For the two directions of crossing.\quad
\tn{b}\,$A_0+k_0B+k_1h/(1+h/h_s)$, with $h=(B-B^*)_+$.
\end{minipage}
\end{table}

\snote{Mode~2 in published absorption spectra of thinner samples}\label{sn:thickness}

Mode~2 lies within the range of two published measurements of the absolute
infrared absorption of FePS$_3$, neither of which reports it. Panda
et~al.~\cite{Panda2026} measured the transmission of a 26-$\mu$m-thick slab between 5
and 150~K, for two orthogonal polarizations, with data reliable from
40~cm$^{-1}$, and Jouanne and Julien~\cite{JouanneJulien1988} the effective
absorption coefficient of 10-$\mu$m-thick slabs at 5 and 300~K, from
50~cm$^{-1}$.
Mode~1 lies below the lower limit of both data sets. To test whether the smaller
thickness explains the absence of mode~2, we computed the effect that a bulk
absorption with the strength measured here would have on those spectra.

\paragraph{Input.}
The oscillator of mode~2 was taken from the fits of Fig.~3 at 1.6 and 20~K, for
the spectra at 5~K, and interpolated to 150~K. We first compared it with the depth
of the band in the transmittance of our crystal (Fig.~1 of the main text): fitting
$\ln\mathcal{T}$ as a smooth polynomial minus $s\,\alpha(\nu)\,d$, with $\alpha$ the
absorption coefficient of the oscillator (background refractive index 2.65) and
$d=298~\mu$m, gives $s=0.84$--0.95 in the well-conditioned cases, so that the
oscillator accounts for the measured band to within 5--16\%. We used $s=0.85$,
the lower end of this range, which makes the test conservative.

\paragraph{Panda et al.}
The transmission spectra of the 26-$\mu$m slab were extracted from the figures of
ref.~\citenum{Panda2026}. They were described as a coherent plane-parallel slab
with a smooth background dielectric function fitted to them (a broad Lorentzian
plus the dispersion of the higher-frequency phonons or, as a deliberately flexible
alternative, a refractive index and a logarithm of the extinction coefficient
quadratic in frequency), to which the oscillator of mode~2 was added. The main
test uses the average of the two polarizations: for any in-plane orientation
$\phi$ of the polarization, $\mathcal{T}(\phi)=\mathcal{T}_a\cos^2\phi+\mathcal{T}_b\sin^2\phi$, so that
the average of two orthogonal polarizations does not depend on the orientation of
the principal axes or on the twin populations of either crystal.

Supplementary Table~\ref{tab:thickness} summarizes the results. With the
background fixed, mode~2 would reduce the averaged transmission by 17\% near
72~cm$^{-1}$ at 5~K and by 25\% near 66~cm$^{-1}$ at 150~K
($\Delta\mathcal{T}\approx0.12$ in both cases), 8--13 times the root-mean-square
deviation of the data from the background (0.009--0.015). For a single
polarization, the reductions are 13--36\%. When the background is refitted with
the mode included, the residuals increase 1.2--2.5 times, and the largest
strength compatible with the data at the 95\% level is 0.07--0.16 of ours
(0.11--0.40 with the flexible background; 0.23--0.32 for a line twice as broad).
Fitted for each polarization separately, the scale of the mode is at most 0.05 in
11 of the 12 combinations of temperature, parameters and polarization. A bulk
absorption with the strength measured in our 298-$\mu$m crystal would therefore be
visible in the 26-$\mu$m slab at 5 and 150~K, in any orientation.

\paragraph{Jouanne and Julien.}
For an incoherent 10-$\mu$m slab and unpolarized light, mode~2 would raise the
effective absorption coefficient by $(0.06$--$0.08)\times10^3$~cm$^{-1}$ near
70~cm$^{-1}$ at 5~K, comparable to the width of the line drawn in the published
figure ($\approx0.09\times10^3$~cm$^{-1}$), and by at most
$0.016\times10^3$~cm$^{-1}$ at 300~K. These spectra do not allow a test.

\paragraph{Absorption near 52~cm$^{-1}$.}
The two samples also differ in the opposite sense. Near 52~cm$^{-1}$, the slab of
Panda et al.\ absorbs $\alpha\approx200$--280~cm$^{-1}$ at 5~K, in a broad feature
listed in their work~\cite{Panda2026}, against $\approx40$--70~cm$^{-1}$ in our
crystal at 1.6~K (from the transmittance of Fig.~1, corrected for reflection
losses), and $\approx250$--290 against $\approx75$--100~cm$^{-1}$ at 140--150~K.
Near 75~cm$^{-1}$, in contrast, the background absorption is comparable: 90 and
153~cm$^{-1}$ in the slab at 5 and 150~K, against 76--112 and $\approx135$~cm$^{-1}$
in our crystal at 1.6 and 140~K.

The test assumes that the absorption of mode~2 is a bulk property, proportional
to the thickness, and does not cover a shift of the mode out of the
50--100~cm$^{-1}$ range in the other sample. The average over polarizations is
exact for domains side by side and approximate for domains stacked along the beam.

\begin{table}[htbp]
\caption{\textbf{Predicted effect of mode~2 on the transmission of the
26-$\mu$m slab of ref.~\citenum{Panda2026}} (Supplementary Note~\ref{sn:thickness}).
Largest reduction of the transmission between 50 and 100~cm$^{-1}$,
$1-\mathcal{T}_\mathrm{with}/\mathcal{T}_\mathrm{without}$, for the oscillator of mode~2 scaled by
$s=0.85$, with the background fixed: for the mode of $\theta=0^\circ$ or of
$\theta=90^\circ$ along either polarization, and for the average of the two
polarizations (with the reduction $\Delta\mathcal{T}$ and its position). rms:
root-mean-square deviation of the averaged data from the background. $s_{95}$:
largest scale of the mode compatible with the averaged data at the 95\% level, with
the background refitted (physical or flexible background).}
\label{tab:thickness}
\centering\small
\setlength{\tabcolsep}{4pt}
\begin{tabular}{c c c c c c c c}
\toprule
Spectra & Mode~2 & \multicolumn{3}{c}{Largest reduction} & rms & \multicolumn{2}{c}{$s_{95}$}\\
\cmidrule(lr){3-5}\cmidrule(lr){7-8}
 & parameters & $\theta=0^\circ$ & $\theta=90^\circ$ & Average & & Physical & Flexible\\
\midrule
5~K & 1.6~K & 25\% & 19\% & 17\% (0.12; 72~cm$^{-1}$) & 0.009 & 0.07 & 0.11\\
5~K & 20~K & 26\% & 18--19\% & 16\% (0.10; 70~cm$^{-1}$) & 0.009 & 0.16 & 0.40\\
150~K & 150~K & 13\% & 36\% & 25\% (0.12; 66~cm$^{-1}$) & 0.015 & 0.16 & 0.12\\
\bottomrule
\end{tabular}
\end{table}

\snote{Oscillator fits and two-component description of mode~2 in field}\label{sn:fits}

Supplementary Figs.~\ref{fig:S1} and \ref{fig:S2} show the real and imaginary parts
of the conductivity of the temperature and field series, with the oscillator fits
from which the parameters of Figs.~3 and 4 of the main text were obtained. In each
spectrum, the two modes are described by damped harmonic oscillators (Eq.~(1) of
Methods), with frequency, linewidth and strength common to $\sigma_1$ and
$\sigma_2$. The weaker absorption that appears near the upper limit of the
measured range at high temperature (main text) is not included in the fits.

In the field series, the line shape of mode~2 was described, field by field,
either by one oscillator or by the sum of two oscillators with a common centre,
one narrow and one broad. Both descriptions were fitted to $\sigma_1$, after
subtraction of the baseline and of the oscillator of mode~1, and to $\sigma_2$,
with a linear background, within $\pm0.35$~THz of the peak of mode~2. The
two-component description was retained if it lowered the BIC,
Eq.~(\ref{eq:bic}), by more than 10 and if each component carried at least 5\% of
the total strength. The two components are a parametrization of the line shape:
only the summed line is used, through its centre, $\nu_2$, its FWHM, $\gamma_2$,
and its total strength, $A_2$, and the uncertainty of $\gamma_2$ was obtained by
Monte Carlo sampling of the fitted parameters with their covariance matrix. At
$\theta=0^\circ$ and 1.6~K, the criterion retains two components from 0 to
4.75~T and one component from 5 to 7~T (Supplementary Table~\ref{tab:field}). It
also retains two components at 140~K in part of the field range, at both
orientations (bold labels in Supplementary Fig.~\ref{fig:S2}), and one component in
all other spectra. Of the two-component fits, only those at $\theta=0^\circ$ and
1.6~K enter Fig.~4 (panels a, c, d and g); the spectral weights of Fig.~4e,f use
the fits with one oscillator per mode and do not depend on this choice.

\begin{figure}[p]
\centering
\includegraphics[width=0.80\textwidth]{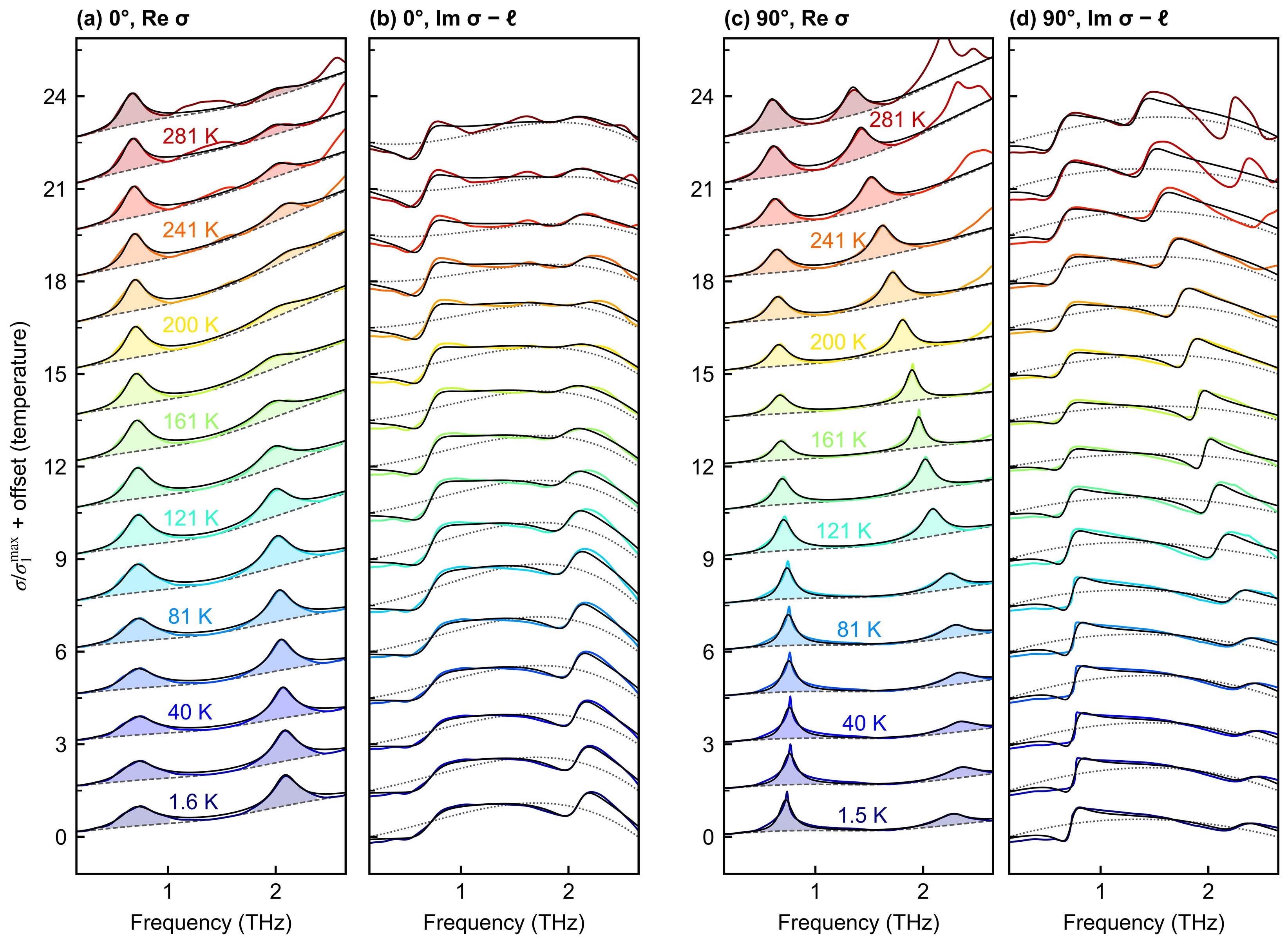}
\caption{\textbf{Oscillator fits of the temperature series at zero field.}
\textbf{a},~\textbf{c} Real part $\sigma_1$ and \textbf{b},~\textbf{d} imaginary
part $\sigma_2-\ell$ of the optical conductivity for $\theta=0^\circ$
(\textbf{a},~\textbf{b}) and $\theta=90^\circ$ (\textbf{c},~\textbf{d}), at every
second measured temperature, from 1.5--1.6~K (bottom) to 301~K (top); every second
spectrum is labelled. Each spectrum is divided by $\sigma_1^\mathrm{max}$, the
maximum of the sum of the two fitted oscillators in $\sigma_1$, and offset
vertically for clarity, as in Fig.~2 of the main text. Coloured lines are the data,
and black lines the fits with one damped harmonic oscillator per mode, obtained
simultaneously from $\sigma_1$ and $\sigma_2$. In \textbf{a} and \textbf{c}, dashed
lines are the baseline $\sigma_b$, to which the fits are added; shaded areas are
the integrals of the fitted oscillator of each mode over $\nu_i\pm0.30$~THz, which
define the spectral weights $W_i$ of Fig.~3g. In \textbf{b} and \textbf{d}, dotted
lines are the smooth non-resonant background $\sigma_2^\mathrm{bg}$, to which the
fits are added; $\ell$, the straight line that joins $\sigma_2^\mathrm{bg}$ at the
ends of the measured range (0.15 and 2.65~THz), is subtracted from all curves for
display only.}
\label{fig:S1}
\end{figure}

\begin{figure}[p]
\centering
\includegraphics[height=0.80\textheight]{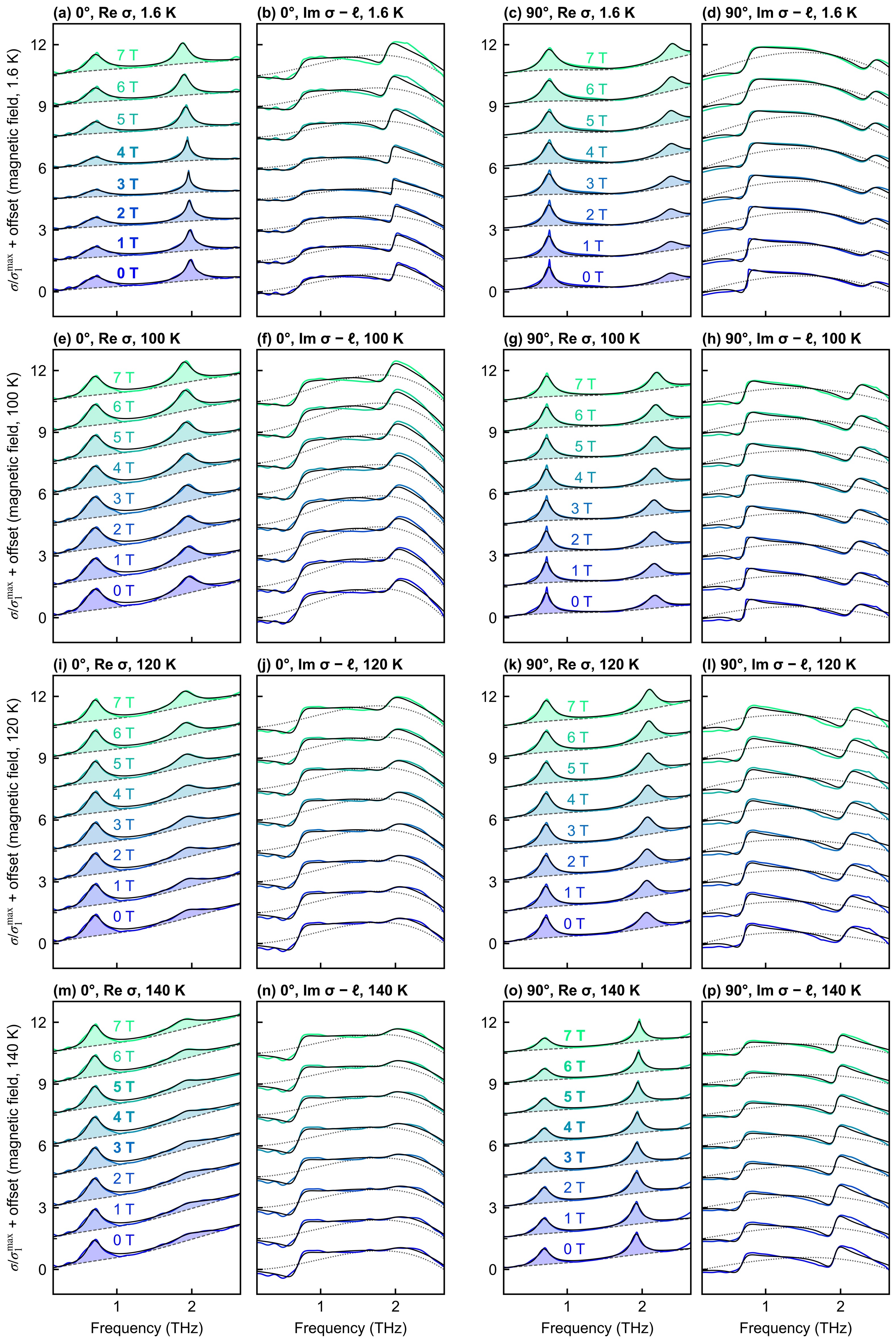}
\caption{\textbf{Oscillator fits of the field series.} Real part $\sigma_1$ and
imaginary part $\sigma_2-\ell$ of the optical conductivity for $\theta=0^\circ$
(first and second columns) and $\theta=90^\circ$ (third and fourth columns), at
1.6~K (\textbf{a}--\textbf{d}), 100~K (\textbf{e}--\textbf{h}), 120~K
(\textbf{i}--\textbf{l}) and 140~K (\textbf{m}--\textbf{p}), from 0~T (bottom) to
7~T (top) in steps of 1~T, with $\mathbf{B}\parallel c^*$. Normalization, offsets,
line styles and the subtraction of $\ell$ are as in Supplementary Fig.~\ref{fig:S1}.
Mode~1 is described by one damped harmonic oscillator, and mode~2 by one
oscillator or, where the field label is in bold, by two oscillators with a common
centre (Supplementary Note~\ref{sn:fits}). Shaded areas in the first and third
columns are the integrals that define the spectral weights of Fig.~4e,f:
$|\sigma_1-\sigma_b|$ minus the fitted oscillator of the other mode, over windows
set for each spectrum.}
\label{fig:S2}
\end{figure}

\snote{Candidate lattice vibrations for the upper THz bands}\label{sn:upperthz}

We reconstructed the harmonic dynamical matrix at $\Gamma$ of the 20-atom antiferromagnetic cell of bulk FePS$_3$ from the finite-displacement forces deposited by Haddadi et al.~\cite{Haddadi2024,HaddadiData2024}. For each of the PBEsol+$U$ and PBEsol+$U$+$V$ calculations, we used 30 displacements in a $2\times2\times2$ supercell, imposed the acoustic sum rule, and diagonalized the mass-weighted matrix. To compare how strongly each vibration bends the Fe--S--Fe bridges, we scaled each displacement pattern to the same maximum atomic displacement of $0.01$~\AA. For each bridge we took half the difference between its angles for opposite displacements, then evaluated the root-mean-square change over all intralayer bridges.

In the same order, the two calculations give 68.3 and $68.2~\mathrm{cm}^{-1}$ for the candidate for $\nu_2$. The corresponding angular changes are $0.208^\circ$ and $0.213^\circ$, compared with only $0.003^\circ$ and $0.013^\circ$ for a nearby branch at 66.8 and $64.3~\mathrm{cm}^{-1}$. This difference favours the first candidate as a vibration capable of modulating exchange through the Fe--S--Fe bridges~\cite{Pestka2024,Wang2025FePX3}. The candidate for $\nu_2'$, at 79.3 and $79.9~\mathrm{cm}^{-1}$, has about 64\% of its mass-weighted eigenvector norm squared on Fe and gives angular changes of just $0.003^\circ$ and $0.001^\circ$. These values characterize the displacement patterns at a common amplitude. The exchange derivative and the evolution of these vibrations with temperature remain undetermined.

Both candidates are even under inversion. The $68.2~\mathrm{cm}^{-1}$ vibration also changes sign under an in-plane translation of the chemical lattice, so a planar stacking fault that preserves this translation is insufficient on its own to activate the mode. The source of its electric-dipole intensity in the measured crystal remains undetermined.

\clearpage
\printbibliography[title={Supplementary References}]

@article{Huang2017,
  author = {Bevin Huang and Genevieve Clark and Efrén Navarro-Moratalla and Dahlia R. Klein and Ran Cheng and Kyle L. Seyler and Ding Zhong and Emma Schmidgall and Michael A. McGuire and David H. Cobden and Wang Yao and Di Xiao and Pablo Jarillo-Herrero and Xiaodong Xu},
  doi = {10.1038/nature22391},
  issn = {0028-0836},
  issue = {7657},
  journal = {Nature},
  month = {6},
  pages = {270-273},
  title = {Layer-dependent ferromagnetism in a van der Waals crystal down to the monolayer limit},
  volume = {546},
  year = {2017}
}

@article{Gong2017,
  author = {Cheng Gong and Lin Li and Zhenglu Li and Huiwen Ji and Alex Stern and Yang Xia and Ting Cao and Wei Bao and Chenzhe Wang and Yuan Wang and Z. Q. Qiu and R. J. Cava and Steven G. Louie and Jing Xia and Xiang Zhang},
  doi = {10.1038/nature22060},
  issn = {0028-0836},
  issue = {7657},
  journal = {Nature},
  month = {6},
  pages = {265-269},
  title = {Discovery of intrinsic ferromagnetism in two-dimensional van der Waals crystals},
  volume = {546},
  year = {2017}
}

@article{Lee2016,
  author = {Jae-Ung Lee and Sungmin Lee and Ji Hoon Ryoo and Soonmin Kang and Tae Yun Kim and Pilkwang Kim and Cheol-Hwan Park and Je-Geun Park and Hyeonsik Cheong},
  doi = {10.1021/acs.nanolett.6b03052},
  issn = {1530-6984},
  issue = {12},
  journal = {Nano Letters},
  month = {12},
  pages = {7433-7438},
  title = {Ising-Type Magnetic Ordering in Atomically Thin FePS$_3$},
  volume = {16},
  year = {2016}
}

@article{Burch2018,
  author = {Burch, Kenneth S. and Mandrus, David and Park, Je-Geun},
  title = {Magnetism in two-dimensional van der {W}aals materials},
  journal = {Nature},
  volume = {563},
  number = {7729},
  pages = {47--52},
  year = {2018},
  doi = {10.1038/s41586-018-0631-z}
}

@article{Gibertini2019,
  author = {Gibertini, M. and Koperski, M. and Morpurgo, A. F. and Novoselov, K. S.},
  title = {Magnetic {2D} materials and heterostructures},
  journal = {Nature Nanotechnology},
  volume = {14},
  number = {5},
  pages = {408--419},
  year = {2019},
  doi = {10.1038/s41565-019-0438-6}
}

@article{GongZhang2019,
  author = {Gong, Cheng and Zhang, Xiang},
  title = {Two-dimensional magnetic crystals and emergent heterostructure devices},
  journal = {Science},
  volume = {363},
  number = {6428},
  pages = {eaav4450},
  year = {2019},
  doi = {10.1126/science.aav4450}
}

@article{Sivadas2018,
  author = {Sivadas, Nikhil and Okamoto, Satoshi and Xu, Xiaodong and Fennie, Craig J. and Xiao, Di},
  title = {Stacking-Dependent Magnetism in Bilayer {CrI$_3$}},
  journal = {Nano Letters},
  volume = {18},
  number = {12},
  pages = {7658--7664},
  year = {2018},
  doi = {10.1021/acs.nanolett.8b03321}
}

@article{Song2019,
  author = {Song, Tiancheng and Fei, Zaiyao and Yankowitz, Matthew and Lin, Zhong and Jiang, Qianni and Hwangbo, Kyle and Zhang, Qi and Sun, Bosong and Taniguchi, Takashi and Watanabe, Kenji and McGuire, Michael A. and Graf, David and Cao, Ting and Chu, Jiun-Haw and Cobden, David H. and Dean, Cory R. and Xiao, Di and Xu, Xiaodong},
  title = {Switching {2D} magnetic states via pressure tuning of layer stacking},
  journal = {Nature Materials},
  volume = {18},
  number = {12},
  pages = {1298--1302},
  year = {2019},
  doi = {10.1038/s41563-019-0505-2}
}

@article{Tan2012,
  author = {Tan, P. H. and Han, W. P. and Zhao, W. J. and Wu, Z. H. and
             Chang, K. and Wang, H. and Wang, Y. F. and Bonini, N. and
             Marzari, N. and Pugno, N. and Savini, G. and Lombardo, A. and
             Ferrari, A. C.},
  title = {The shear mode of multilayer graphene},
  journal = {Nature Materials},
  volume = {11},
  pages = {294--300},
  year = {2012},
  doi = {10.1038/nmat3245}
}

@article{ZhangX2013,
  author = {X. Zhang and W. P. Han and J. B. Wu and S. Milana and Y. Lu and Q. Q. Li and A. C. Ferrari and P. H. Tan},
  doi = {10.1103/PhysRevB.87.115413},
  issn = {1098-0121},
  issue = {11},
  journal = {Physical Review B},
  month = {3},
  pages = {115413},
  title = {Raman spectroscopy of shear and layer breathing modes in multilayer MoS$_2$},
  volume = {87},
  year = {2013}
}

@article{Liang2017,
  author = {Liang, Liangbo and Zhang, Jun and Sumpter, Bobby G. and
             Tan, Qing-Hai and Tan, Ping-Heng and Meunier, Vincent},
  title = {Low-frequency shear and layer-breathing modes in {Raman}
             scattering of two-dimensional materials},
  journal = {ACS Nano},
  volume = {11},
  number = {12},
  pages = {11777--11802},
  year = {2017},
  doi = {10.1021/acsnano.7b06551}
}

@article{Brec1986,
  author = {Brec, R.},
  title = {Review on structural and chemical properties of transition metal phosphorous trisulfides {MPS$_3$}},
  journal = {Solid State Ionics},
  volume = {22},
  number = {1},
  pages = {3--30},
  year = {1986},
  doi = {10.1016/0167-2738(86)90055-X}
}

@article{Susner2017,
  author = {Susner, Michael A. and Chyasnavichyus, Marius and McGuire, Michael A. and Ganesh, Panchapakesan and Maksymovych, Petro},
  title = {Metal Thio- and Selenophosphates as Multifunctional van der {W}aals Layered Materials},
  journal = {Advanced Materials},
  volume = {29},
  number = {38},
  pages = {1602852},
  year = {2017},
  doi = {10.1002/adma.201602852}
}

@article{Ouvrard1985,
  author = {Ouvrard, G. and Brec, R. and Rouxel, J.},
  title = {Structural determination of some {MPS$_3$} layered phases ({M} = {Mn}, {Fe}, {Co}, {Ni} and {Cd})},
  journal = {Materials Research Bulletin},
  volume = {20},
  number = {10},
  pages = {1181--1189},
  year = {1985},
  doi = {10.1016/0025-5408(85)90092-3}
}

@article{Kurosawa1983,
  author = {Kurosawa, Ko and Saito, Shozo and Yamaguchi, Yasuo},
  title = {Neutron Diffraction Study on {MnPS$_3$} and {FePS$_3$}},
  journal = {Journal of the Physical Society of Japan},
  volume = {52},
  number = {11},
  pages = {3919--3926},
  year = {1983},
  doi = {10.1143/JPSJ.52.3919}
}

@article{Lancon2016,
  author = {D. Lançon and H. C. Walker and E. Ressouche and B. Ouladdiaf and K. C. Rule and G. J. McIntyre and T. J. Hicks and H. M. Rønnow and A. R. Wildes},
  doi = {10.1103/PhysRevB.94.214407},
  issn = {2469-9950},
  issue = {21},
  journal = {Physical Review B},
  month = {12},
  pages = {214407},
  title = {Magnetic structure and magnon dynamics of the quasi-two-dimensional antiferromagnet FePS$_3$},
  volume = {94},
  year = {2016}
}

@article{Jernberg1984,
  author = {Jernberg, P. and Bjarman, S. and W\"appling, R.},
  title = {{FePS$_3$}: A first order phase transition in a {2D} {I}sing
             antiferromagnet},
  journal = {Journal of Magnetism and Magnetic Materials},
  volume = {46},
  pages = {178--190},
  year = {1984},
  doi = {10.1016/0304-8853(84)90355-X}
}

@article{Joy1992,
  author = {Joy, P. A. and Vasudevan, S.},
  title = {Magnetism in the layered transition-metal thiophosphates
             {$M$PS$_3$} ({$M$} = {M}n, {F}e, and {N}i)},
  journal = {Physical Review B},
  volume = {46},
  pages = {5425--5433},
  year = {1992},
  doi = {10.1103/PhysRevB.46.5425}
}

@article{Rule2007,
  author = {Rule, K. C. and McIntyre, G. J. and Kennedy, S. J. and Hicks, T. J.},
  title = {Single-crystal and powder neutron diffraction experiments on {FePS$_3$}: Search for the magnetic structure},
  journal = {Physical Review B},
  volume = {76},
  number = {13},
  pages = {134402},
  year = {2007},
  doi = {10.1103/PhysRevB.76.134402}
}

@article{Liu2021,
  author = {Sheng Liu and Andrés Granados del Águila and Dhiman Bhowmick and Chee Kwan Gan and T. Thu Ha Do and M. A. Prosnikov and David Sedmidubský and Zdenek Sofer and Peter C. M. Christianen and Pinaki Sengupta and Qihua Xiong},
  doi = {10.1103/PhysRevLett.127.097401},
  issn = {0031-9007},
  issue = {9},
  journal = {Physical Review Letters},
  month = {8},
  pages = {097401},
  title = {Direct Observation of Magnon-Phonon Strong Coupling in Two-Dimensional Antiferromagnet at High Magnetic Fields},
  volume = {127},
  year = {2021}
}

@misc{ZhangQ2021,
  title = {Coherent strong-coupling of terahertz magnons and phonons in a Van der Waals antiferromagnetic insulator},
  author = {Qi Zhang and Mykhaylo Ozerov and Emil Vinas Boström and Jun Cui and Nishchay Suri and Qianni Jiang and Chong Wang and Fangliang Wu and Kyle Hwangbo and Jiun-Haw Chu and Di Xiao and Angel Rubio and Xiaodong Xu},
  year = {2021},
  eprint = {2108.11619},
  archiveprefix = {arXiv},
  primaryclass = {cond-mat.mtrl-sci},
  url = {https://arxiv.org/abs/2108.11619}
}

@article{Wang2016,
  author = {Xingzhi Wang and Kezhao Du and Yu Yang Fredrik Liu and Peng Hu and Jun Zhang and Qing Zhang and Man Hon Samuel Owen and Xin Lu and Chee Kwan Gan and Pinaki Sengupta and Christian Kloc and Qihua Xiong},
  doi = {10.1088/2053-1583/3/3/031009},
  issn = {2053-1583},
  issue = {3},
  journal = {2D Materials},
  month = {8},
  pages = {031009},
  title = {Raman spectroscopy of atomically thin two-dimensional magnetic iron phosphorus trisulfide (FePS$_3$) crystals},
  volume = {3},
  year = {2016}
}

@article{Murayama2016,
  author = {Murayama, C. and Okabe, M. and Urushihara, D. and Asaka, T. and
             Fukuda, K. and Isobe, M. and Yamamoto, K. and Matsushita, Y.},
  title = {Crystallographic features related to a van der {W}aals coupling
             in the layered chalcogenide {FePS$_3$}},
  journal = {Journal of Applied Physics},
  volume = {120},
  pages = {142114},
  year = {2016},
  doi = {10.1063/1.4961712}
}

@article{ZhangQ2021LD,
  author = {Zhang, Qi and Hwangbo, Kyle and Wang, Chong and Jiang, Qianni and Chu, Jiun-Haw and Wen, Haidan and Xiao, Di and Xu, Xiaodong},
  title = {Observation of Giant Optical Linear Dichroism in a Zigzag Antiferromagnet {FePS$_3$}},
  journal = {Nano Letters},
  volume = {21},
  number = {16},
  pages = {6938--6945},
  year = {2021},
  doi = {10.1021/acs.nanolett.1c02188}
}

@article{Haines2018,
  author = {Haines, C. R. S. and Coak, M. J. and Wildes, A. R. and
             Lampronti, G. I. and Liu, C. and Nahai-Williamson, P. and
             Hamidov, H. and Daisenberger, D. and Saxena, S. S.},
  title = {Pressure-induced electronic and structural phase evolution in the
             van der {W}aals compound {FePS$_3$}},
  journal = {Physical Review Letters},
  volume = {121},
  pages = {266801},
  year = {2018},
  doi = {10.1103/PhysRevLett.121.266801}
}

@article{Coak2021,
  author = {Coak, Matthew J. and Jarvis, David M. and Hamidov, Hayrullo and Wildes, Andrew R. and Paddison, Joseph A. M. and Liu, Cheng and Haines, Charles R. S. and Dang, Ngoc T. and Kichanov, Sergey E. and Savenko, Boris N. and Lee, Sungmin and Kratochv{\'\i}lov{\'a}, Marie and Klotz, Stefan and Hansen, Thomas C. and Kozlenko, Denis P. and Park, Je-Geun and Saxena, Siddharth S.},
  title = {Emergent Magnetic Phases in Pressure-Tuned van der {W}aals Antiferromagnet {FePS$_3$}},
  journal = {Physical Review X},
  volume = {11},
  number = {1},
  pages = {011024},
  year = {2021},
  doi = {10.1103/PhysRevX.11.011024}
}

@article{Dhakal2024,
  author = {Ramesh Dhakal and Samuel Griffith and Stephen M. Winter},
  doi = {10.1038/s41535-024-00675-x},
  issn = {2397-4648},
  issue = {1},
  journal = {npj Quantum Materials},
  month = {8},
  pages = {64},
  title = {Hybrid spin-orbit exciton-magnon excitations in FePS3},
  volume = {9},
  year = {2024}
}

@article{Sekine1990,
  author = {T. Sekine and M. Jouanne and C. Julien and M. Balkanski},
  doi = {10.1103/PhysRevB.42.8382},
  issn = {0163-1829},
  issue = {13},
  journal = {Physical Review B},
  month = {11},
  pages = {8382-8393},
  title = {Light-scattering study of dynamical behavior of antiferromagnetic spins in the layered magnetic semiconductor FePS$_3$},
  volume = {42},
  year = {1990}
}

@article{McCreary2020,
  author = {Amber McCreary and Jeffrey R. Simpson and Thuc T. Mai and Robert D. McMichael and Jason E. Douglas and Nicholas Butch and Cindi Dennis and Rolando Valdés Aguilar and Angela R. Hight Walker},
  doi = {10.1103/PhysRevB.101.064416},
  issn = {2469-9950},
  issue = {6},
  journal = {Physical Review B},
  month = {2},
  pages = {064416},
  title = {Quasi-two-dimensional magnon identification in antiferromagnetic FePS$_3$ via magneto-Raman spectroscopy},
  volume = {101},
  year = {2020}
}

@article{Vaclavkova2021,
  author = {D. Vaclavkova and M. Palit and J. Wyzula and S. Ghosh and A. Delhomme and S. Maity and P. Kapuscinski and A. Ghosh and M. Veis and M. Grzeszczyk and C. Faugeras and M. Orlita and S. Datta and M. Potemski},
  doi = {10.1103/PhysRevB.104.134437},
  issn = {2469-9950},
  issue = {13},
  journal = {Physical Review B},
  month = {10},
  pages = {134437},
  title = {Magnon polarons in the van der Waals antiferromagnet FePS$_3$},
  volume = {104},
  year = {2021}
}

@article{Ghosh2021,
  author = {Anudeepa Ghosh and Mainak Palit and Sujan Maity and Vivek Dwij and Sumesh Rana and Subhadeep Datta},
  doi = {10.1103/PhysRevB.103.064431},
  issn = {2469-9950},
  issue = {6},
  journal = {Physical Review B},
  month = {2},
  pages = {064431},
  title = {Spin-phonon coupling and magnon scattering in few-layer antiferromagnetic FePS$_3$},
  volume = {103},
  year = {2021}
}

@article{Scagliotti1987,
  author = {Scagliotti, M. and Jouanne, M. and Balkanski, M. and Ouvrard, G. and Benedek, G.},
  title = {Raman scattering in antiferromagnetic {FePS$_3$} and {FePSe$_3$} crystals},
  journal = {Physical Review B},
  volume = {35},
  number = {13},
  pages = {7097--7104},
  year = {1987},
  doi = {10.1103/PhysRevB.35.7097}
}

@article{Ergecen2023,
  author = {Emre Ergeçen and Batyr Ilyas and Junghyun Kim and Jaena Park and Mehmet Burak Yilmaz and Tianchuang Luo and Di Xiao and Satoshi Okamoto and Je-Geun Park and Nuh Gedik},
  doi = {10.1073/pnas.2208968120},
  issn = {0027-8424},
  issue = {12},
  journal = {Proceedings of the National Academy of Sciences},
  month = {3},
  title = {Coherent detection of hidden spin–lattice coupling in a van der Waals antiferromagnet},
  volume = {120},
  year = {2023}
}

@article{Mertens2023,
  author = {Fabian Mertens and David Mönkebüscher and Umut Parlak and Carla Boix‐Constant and Samuel Mañas‐Valero and Margherita Matzer and Rajdeep Adhikari and Alberta Bonanni and Eugenio Coronado and Alexandra M. Kalashnikova and Davide Bossini and Mirko Cinchetti},
  doi = {10.1002/adma.202208355},
  issn = {0935-9648},
  issue = {6},
  journal = {Advanced Materials},
  month = {2},
  title = {Ultrafast Coherent THz Lattice Dynamics Coupled to Spins in the van der Waals Antiferromagnet FePS$_3$},
  volume = {35},
  year = {2023}
}

@article{ZhangXX2021,
  author = {Zhang, Xiao-Xiao and Jiang, Shengwei and Lee, Jinhwan and Lee, Changgu and Mak, Kin Fai and Shan, Jie},
  title = {Spin Dynamics Slowdown near the Antiferromagnetic Critical Point in Atomically Thin {FePS$_3$}},
  journal = {Nano Letters},
  volume = {21},
  number = {12},
  pages = {5045--5052},
  year = {2021},
  doi = {10.1021/acs.nanolett.1c00870}
}

@article{Zhou2022,
  author = {Faran Zhou and Kyle Hwangbo and Qi Zhang and Chong Wang and Lingnan Shen and Jiawei Zhang and Qianni Jiang and Alfred Zong and Yifan Su and Marc Zajac and Youngjun Ahn and Donald A. Walko and Richard D. Schaller and Jiun-Haw Chu and Nuh Gedik and Xiaodong Xu and Di Xiao and Haidan Wen},
  doi = {10.1038/s41467-022-34376-5},
  issn = {2041-1723},
  issue = {1},
  journal = {Nature Communications},
  month = {11},
  pages = {6598},
  title = {Dynamical criticality of spin-shear coupling in van der Waals antiferromagnets},
  volume = {13},
  year = {2022}
}

@article{Zong2023,
  author = {Alfred Zong and Qi Zhang and Faran Zhou and Yifan Su and Kyle Hwangbo and Xiaozhe Shen and Qianni Jiang and Haihua Liu and Thomas E. Gage and Donald A. Walko and Michael E. Kozina and Duan Luo and Alexander H. Reid and Jie Yang and Suji Park and Saul H. Lapidus and Jiun-Haw Chu and Ilke Arslan and Xijie Wang and Di Xiao and Xiaodong Xu and Nuh Gedik and Haidan Wen},
  doi = {10.1038/s41586-023-06279-y},
  issn = {0028-0836},
  issue = {7976},
  journal = {Nature},
  month = {8},
  pages = {988-993},
  title = {Spin-mediated shear oscillators in a van der Waals antiferromagnet},
  volume = {620},
  year = {2023}
}

@article{Ilyas2024,
  author = {Batyr Ilyas and Tianchuang Luo and Alexander von Hoegen and Emil Viñas Boström and Zhuquan Zhang and Jaena Park and Junghyun Kim and Je-Geun Park and Keith A. Nelson and Angel Rubio and Nuh Gedik},
  doi = {10.1038/s41586-024-08226-x},
  issn = {0028-0836},
  issue = {8043},
  journal = {Nature},
  month = {12},
  pages = {609-614},
  title = {Terahertz field-induced metastable magnetization near criticality in FePS$_3$},
  volume = {636},
  year = {2024}
}

@article{Luo2025,
  author = {Tianchuang Luo and Honglie Ning and Batyr Ilyas and Alexander von Hoegen and Emil Viñas Boström and Jaena Park and Junghyun Kim and Je-Geun Park and Dominik M. Juraschek and Angel Rubio and Nuh Gedik},
  doi = {10.1038/s41467-025-62091-4},
  issn = {2041-1723},
  issue = {1},
  journal = {Nature Communications},
  month = {7},
  pages = {6863},
  title = {Terahertz control of linear and nonlinear Magno-phononics},
  volume = {16},
  year = {2025}
}

@article{Li2026,
  author = {Liangyue Li and Na Wu and Zhengwang Lin and Zefen Li and Lixin Liu and Emil Vinas Boström and Yuan Wan and Xinbo Wang and Jianlin Luo and Fucai Liu and Angel Rubio and Qi Zhang},
  doi = {10.1038/s41467-026-77233-5},
  issn = {2041-1723},
  issue = {1},
  journal = {Nature Communications},
  month = {9},
  pages = {9507},
  title = {Coherent terahertz magnon–phonon three-wave mixing in a layered antiferromagnet},
  volume = {17},
  year = {2026}
}

@article{Wyzula2022,
  author = {Jan Wyzula and Ivan Mohelský and Diana Václavková and Piotr Kapuscinski and Martin Veis and Clément Faugeras and Marek Potemski and Mike E. Zhitomirsky and Milan Orlita},
  doi = {10.1021/acs.nanolett.2c04111},
  issn = {1530-6984},
  issue = {23},
  journal = {Nano Letters},
  month = {12},
  pages = {9741-9747},
  title = {High-Angular Momentum Excitations in Collinear Antiferromagnet FePS 3},
  volume = {22},
  year = {2022}
}

@article{LeMardel2024,
  author = {F. Le Mardelé and I. Mohelsky and D. Jana and A. Pawbake and J. Dzian and W.-L. Lee and K. Raju and R. Sankar and C. Faugeras and M. Potemski and M. E. Zhitomirsky and M. Orlita},
  doi = {10.1103/PhysRevB.110.174414},
  issn = {2469-9950},
  issue = {17},
  journal = {Physical Review B},
  month = {11},
  pages = {174414},
  title = {Tuning terahertz magnons in a mixed van der Waals antiferromagnet},
  volume = {110},
  year = {2024}
}

@article{JouanneJulien1988,
  author = {M. Jouanne and C. Julien},
  doi = {10.1063/1.341402},
  issn = {0021-8979},
  issue = {7},
  journal = {Journal of Applied Physics},
  month = {10},
  pages = {3637-3640},
  title = {Fourier transform infrared spectroscopy studies at low temperatures of MPX3 layered compounds},
  volume = {64},
  year = {1988}
}

@article{Panda2026,
  author = {Kartik Panda and Itzik Kapon and Dumitru Dumcenco and Dirk van der Marel and Alexey B. Kuzmenko and Nimrod Bachar},
  doi = {10.1103/fwb8-fwjb},
  issn = {2469-9950},
  issue = {23},
  journal = {Physical Review B},
  month = {6},
  pages = {235124},
  title = {Coupling of magnetic and lattice collective excitations in the two-dimensional van der Waals antiferromagnet   FePS 3  },
  volume = {113},
  year = {2026}
}

@article{Grischkowsky1990,
  author = {Grischkowsky, D. and Keiding, S{\o}ren and van Exter, Martin and Fattinger, Ch.},
  title = {Far-infrared time-domain spectroscopy with terahertz beams of dielectrics and semiconductors},
  journal = {Journal of the Optical Society of America B},
  volume = {7},
  number = {10},
  pages = {2006--2015},
  year = {1990},
  doi = {10.1364/JOSAB.7.002006}
}

@article{Lloyd-Hughes2012-es,
  author = {James Lloyd-Hughes and Tae-In Jeon},
  doi = {10.1007/s10762-012-9905-y},
  issn = {1866-6892},
  issue = {9},
  journal = {Journal of Infrared, Millimeter, and Terahertz Waves},
  month = {9},
  pages = {871-925},
  title = {A Review of the Terahertz Conductivity of Bulk and Nano-Materials},
  volume = {33},
  year = {2012}
}

@article{marulanda2026colossal,
  title = {{Colossal terahertz magnetoresistance from magnetic polarons in EuZn${}_{2}$P${}_{2}$}},
  author = {Marulanda, E. and Dutra, M. and Kawahala, N. M. and Stefanato, E. D. and Vasques, G. G. and Munevar, J. and Avila, M. A. and Hernandez, F. G. G.},
  journal = {Phys. Rev. Lett.},
  pages = {},
  year = {2026},
  month = nov,
  publisher = {American Physical Society},
  doi = {10.1103/dlh5-6ffv},
  url = {https://link.aps.org/doi/10.1103/dlh5-6ffv}
}

@article{Kawahala2026,
  author = {Nicolas M. Kawahala and Rafael L. Sabainsk and Esteban Marulanda and Rafael S. Freitas and Felix G. G. Hernandez},
  doi = {10.1103/m693-z3kk},
  issn = {2643-1564},
  issue = {4},
  journal = {Physical Review Research},
  month = {11},
  pages = {043114},
  title = {Terahertz lattice imprint of a hidden multipolar ordering transition in    Gd2Ru2O7},
  volume = {8},
  year = {2026}
}

@article{Wildes2020,
  author = {A. R. Wildes and D. Lançon and M. K. Chan and F. Weickert and N. Harrison and V. Simonet and M. E. Zhitomirsky and M. V. Gvozdikova and T. Ziman and H. M. Rønnow},
  doi = {10.1103/PhysRevB.101.024415},
  issn = {2469-9950},
  issue = {2},
  journal = {Physical Review B},
  month = {1},
  pages = {024415},
  title = {High field magnetization of FePS$_3$},
  volume = {101},
  year = {2020}
}

@article{Zhou2026,
  author = {Faran Zhou and Kyle Hwangbo and Sung Soo Ha and Xiao‐Wei Zhang and Sae Hwan Chun and Jaeku Park and Intae Eom and Qianni Jiang and Zekai Yang and Marc Zajac and Sungwon Kim and Sungwook Choi and Zhaodong Chu and Kyoung Hun Oh and Yifan Su and Alfred Zong and Elton J. G. Santos and Ting Cao and Jiun‐Haw Chu and Stephan O. Hruszkewycz and Nuh Gedik and Di Xiao and Hyunjung Kim and Xiaodong Xu and Haidan Wen},
  doi = {10.1002/adma.202522280},
  issn = {0935-9648},
  issue = {14},
  journal = {Advanced Materials},
  month = {3},
  title = {Atomically Resolved Acoustic Dynamics Coupled with Magnetic Order in a Van der Waals Antiferromagnet},
  volume = {38},
  year = {2026}
}

@article{Nguyen2025,
  author = {Nguyen, Manh Hong and Park, Giung and Park, Je-Geun and Cheong, Hyeonsik},
  title = {Interlayer interaction and {D}avydov splitting in antiferromagnetic few-layer {NiPS$_3$}},
  journal = {npj Quantum Materials},
  volume = {10},
  number = {1},
  pages = {65},
  year = {2025},
  doi = {10.1038/s41535-025-00788-x}
}

@article{Kant2012,
  author = {Kant, Ch. and Schmidt, M. and Wang, Zhe and Mayr, F. and Tsurkan, V. and Deisenhofer, J. and Loidl, A.},
  title = {Universal Exchange-Driven Phonon Splitting in Antiferromagnets},
  journal = {Physical Review Letters},
  volume = {108},
  number = {17},
  pages = {177203},
  year = {2012},
  doi = {10.1103/PhysRevLett.108.177203}
}

@article{Baltensperger1968,
  author = {Baltensperger, W. and Helman, J. S.},
  title = {Influence of magnetic order in insulators on the optical phonon
             frequency},
  journal = {Helvetica Physica Acta},
  volume = {41},
  pages = {668--673},
  year = {1968}
}

@article{Granado1999,
  author = {Granado, E. and Garc{\'i}a, A. and Sanjurjo, J. A. and
             Rettori, C. and Torriani, I. and Prado, F. and S{\'a}nchez, R. D.
             and Caneiro, A. and Oseroff, S. B.},
  title = {Magnetic ordering effects in the {R}aman spectra of
             {La$_{1-x}$Mn$_{1-x}$O$_3$}},
  journal = {Physical Review B},
  volume = {60},
  pages = {11879--11882},
  year = {1999},
  doi = {10.1103/PhysRevB.60.11879}
}

@article{Sushkov2005,
  author = {Sushkov, A. B. and Tchernyshyov, O. and {Ratcliff II}, W. and Cheong, S.-W. and Drew, H. D.},
  title = {Probing Spin Correlations with Phonons in the Strongly Frustrated Magnet {ZnCr$_2$O$_4$}},
  journal = {Physical Review Letters},
  volume = {94},
  number = {13},
  pages = {137202},
  year = {2005},
  doi = {10.1103/PhysRevLett.94.137202}
}

@article{GonzeLee1997,
  author = {Gonze, Xavier and Lee, Changyol},
  title = {Dynamical matrices, {B}orn effective charges, dielectric permittivity tensors, and interatomic force constants from density-functional perturbation theory},
  journal = {Physical Review B},
  volume = {55},
  number = {16},
  pages = {10355--10368},
  year = {1997},
  doi = {10.1103/PhysRevB.55.10355}
}

@article{Baroni2001,
  author = {Baroni, Stefano and de Gironcoli, Stefano and Dal Corso, Andrea
             and Giannozzi, Paolo},
  title = {Phonons and related crystal properties from density-functional
             perturbation theory},
  journal = {Reviews of Modern Physics},
  volume = {73},
  pages = {515--562},
  year = {2001},
  doi = {10.1103/RevModPhys.73.515}
}

@article{Chen2024,
  author = {Lebing Chen and Xiaokun Teng and Ding Hu and Feng Ye and Garrett E. Granroth and Ming Yi and Jae-Ho Chung and Robert J. Birgeneau and Pengcheng Dai},
  doi = {10.1038/s41535-024-00651-5},
  issn = {2397-4648},
  issue = {1},
  journal = {npj Quantum Materials},
  month = {5},
  pages = {40},
  title = {Thermal evolution of spin excitations in honeycomb Ising antiferromagnetic FePSe3},
  volume = {9},
  year = {2024}
}

@article{HohenbergHalperin1977,
  author = {Hohenberg, P. C. and Halperin, B. I.},
  title = {Theory of dynamic critical phenomena},
  journal = {Reviews of Modern Physics},
  volume = {49},
  number = {3},
  pages = {435--479},
  year = {1977},
  doi = {10.1103/RevModPhys.49.435}
}

@article{kawahala_shaping_2025,
  title = {Shaping terahertz waves using anisotropic shear modes in a van der {Waals} mineral},
  volume = {9},
  copyright = {2025 The Author(s)},
  issn = {2397-7132},
  url = {https://www.nature.com/articles/s41699-025-00540-w},
  doi = {10.1038/s41699-025-00540-w},
  number = {1},
  journal = {npj 2D Materials and Applications},
  author = {Kawahala, Nicolas M. and Matos, Daniel A. and de Oliveira, Raphaela and Longuinhos, Raphael and Ribeiro-Soares, Jenaina and Barcelos, Ingrid D. and Hernandez, Felix G. G.},
  month = feb,
  year = {2025},
  pages = {16}
}

@article{BarkerSievers1975,
  author = {Barker, Jr., A. S. and Sievers, A. J.},
  title = {Optical studies of the vibrational properties of disordered solids},
  journal = {Reviews of Modern Physics},
  volume = {47},
  number = {S2},
  pages = {S1--S179},
  year = {1975},
  doi = {10.1103/RevModPhys.47.S1.2}
}

@article{Haddadi2024,
  author = {Haddadi, Fatemeh and Linscott, Edward and Timrov, Iurii and Marzari, Nicola and Gibertini, Marco},
  title = {On-site and intersite Hubbard corrections in magnetic monolayers: The case of {FePS$_3$} and {CrI$_3$}},
  journal = {Physical Review Materials},
  year = {2024},
  volume = {8},
  pages = {014007},
  doi = {10.1103/PhysRevMaterials.8.014007}
}

@misc{HaddadiData2024,
  author = {Haddadi, Fatemeh and Linscott, Edward and Timrov, Iurii and Marzari, Nicola and Gibertini, Marco},
  title = {On-site and inter-site Hubbard corrections in magnetic monolayers: The case of {FePS$_3$} and {CrI$_3$}},
  publisher = {Materials Cloud},
  year = {2024},
  doi = {10.24435/materialscloud:ez-6k}
}

@article{Pestka2024,
  author = {Pestka, Benjamin and Strasdas, Jeff and Bihlmayer, Gustav and Budniak, Adam Krzysztof and Liebmann, Marcus and Leuth, Niklas and Boban, Honey and Feyer, Vitaliy and Cojocariu, Iulia and Baranowski, Daniel and Mearini, Simone and Amouyal, Yaron and Waldecker, Lutz and Beschoten, Bernd and Stampfer, Christoph and Plucinski, Lukasz and Lifshitz, Efrat and Kratzer, Peter and Morgenstern, Markus},
  title = {Identifying Band Structure Changes of {FePS$_3$} across the Antiferromagnetic Phase Transition},
  journal = {ACS Nano},
  year = {2024},
  volume = {18},
  pages = {32924--32931},
  doi = {10.1021/acsnano.4c12520}
}

@article{Wang2025FePX3,
  author = {Wang, Qingyang and Niu, Mengmeng and Zhou, Weikang and Ma, Yicheng and Huang, Chun and Yang, Gege and Shao, Yan and Wu, Xu and Wang, Cong and Ji, Wei and Wang, Yeliang and Qiao, Jingsi},
  title = {Magnetic and Multiferroic Properties of Two-Dimensional {FePX$_3$} and {CuFeP$_2$X$_6$} ({X = S, Se, and Te})},
  journal = {ACS Applied Electronic Materials},
  year = {2025},
  volume = {7},
  pages = {660--666},
  doi = {10.1021/acsaelm.4c01547}
}

@article{Dias2026,
  author = {Laura O. Dias and Eduardo D. Stefanato and Nicolas M. Kawahala and Felix G. G. Hernandez},
  doi = {10.1007/s13538-025-01973-w},
  issn = {0103-9733},
  issue = {1},
  journal = {Brazilian Journal of Physics},
  month = {2},
  pages = {35},
  title = {Optimizing Aperture Geometry in THz-TDS for Accurate Spectroscopy of Quantum Materials},
  volume = {56},
  year = {2026}
}

@article{Marulanda2025,
  author = {Esteban Marulanda and Fernanda L. Costa and Nicolas M. Kawahala and Felix G. G. Hernandez},
  doi = {10.1007/s10762-025-01092-2},
  issn = {1866-6892},
  issue = {10},
  journal = {Journal of Infrared, Millimeter, and Terahertz Waves},
  month = {10},
  pages = {75},
  title = {Windowing in Terahertz Time-Domain Spectroscopy: Resolving Resonances in Thin-Film Samples},
  volume = {46},
  year = {2025}
}

@techreport{EilersBoelens2005,
  author = {Eilers, Paul H. C. and Boelens, Hans F. M.},
  title = {Baseline Correction with Asymmetric Least Squares Smoothing},
  institution = {Leiden University Medical Centre},
  year = {2005}
}

@article{Barrett1952,
  author = {Barrett, John H.},
  title = {Dielectric Constant in Perovskite Type Crystals},
  journal = {Physical Review},
  volume = {86},
  number = {1},
  pages = {118--120},
  year = {1952},
  doi = {10.1103/PhysRev.86.118}
}

@article{Klemens1966,
  author = {Klemens, P. G.},
  title = {Anharmonic Decay of Optical Phonons},
  journal = {Physical Review},
  volume = {148},
  number = {2},
  pages = {845--848},
  year = {1966},
  doi = {10.1103/PhysRev.148.845}
}

@article{Balkanski1983,
  author = {M. Balkanski and R. F. Wallis and E. Haro},
  doi = {10.1103/PhysRevB.28.1928},
  issn = {0163-1829},
  issue = {4},
  journal = {Physical Review B},
  month = {8},
  pages = {1928-1934},
  title = {Anharmonic effects in light scattering due to optical phonons in silicon},
  volume = {28},
  year = {1983}
}

@article{MenendezCardona1984,
  author = {Men{\'e}ndez, Jos{\'e} and Cardona, Manuel},
  title = {Temperature dependence of the first-order {R}aman scattering by phonons in {Si}, {Ge}, and $\alpha$-{Sn}: Anharmonic effects},
  journal = {Physical Review B},
  volume = {29},
  number = {4},
  pages = {2051--2059},
  year = {1984},
  doi = {10.1103/PhysRevB.29.2051}
}

@article{Houtappel1950,
  author = {Houtappel, R. M. F.},
  title = {Order-disorder in hexagonal lattices},
  journal = {Physica},
  volume = {16},
  number = {5},
  pages = {425--455},
  year = {1950},
  doi = {10.1016/0031-8914(50)90130-3}
}

@article{Potts1952,
  author = {Potts, R. B.},
  title = {Spontaneous Magnetization of a Triangular {I}sing Lattice},
  journal = {Physical Review},
  volume = {88},
  number = {2},
  pages = {352},
  year = {1952},
  doi = {10.1103/PhysRev.88.352}
}

@book{Baxter1982,
  author = {Baxter, Rodney J.},
  title = {Exactly Solved Models in Statistical Mechanics},
  publisher = {Academic Press},
  address = {London},
  year = {1982}
}

@article{Yang1952,
  author = {Yang, C. N.},
  title = {The Spontaneous Magnetization of a Two-Dimensional {I}sing Model},
  journal = {Physical Review},
  volume = {85},
  number = {5},
  pages = {808--816},
  year = {1952},
  doi = {10.1103/PhysRev.85.808}
}

@article{Onsager1944,
  author = {Onsager, Lars},
  title = {Crystal Statistics. {I}. {A} Two-Dimensional Model with an Order-Disorder Transition},
  journal = {Physical Review},
  volume = {65},
  number = {3--4},
  pages = {117--149},
  year = {1944},
  doi = {10.1103/PhysRev.65.117}
}

@article{Schwarz1978,
  author = {Schwarz, Gideon},
  title = {Estimating the Dimension of a Model},
  journal = {The Annals of Statistics},
  volume = {6},
  number = {2},
  pages = {461--464},
  year = {1978},
  doi = {10.1214/aos/1176344136}
}

@article{KassRaftery1995,
  author = {Kass, Robert E. and Raftery, Adrian E.},
  title = {Bayes Factors},
  journal = {Journal of the American Statistical Association},
  volume = {90},
  number = {430},
  pages = {773--795},
  year = {1995},
  doi = {10.1080/01621459.1995.10476572}
}

@book{LandauLifshitz1980,
  author = {Landau, L. D. and Lifshitz, E. M.},
  title = {Statistical Physics, Part 1},
  edition = {3rd},
  publisher = {Pergamon},
  address = {Oxford},
  year = {1980}
}

@book{ChaikinLubensky1995,
  author = {Chaikin, P. M. and Lubensky, T. C.},
  title = {Principles of Condensed Matter Physics},
  publisher = {Cambridge University Press},
  address = {Cambridge},
  year = {1995}
}

@article{Rehwald1973,
  author = {Rehwald, W.},
  title = {The study of structural phase transitions by means of ultrasonic experiments},
  journal = {Advances in Physics},
  volume = {22},
  number = {6},
  pages = {721--755},
  year = {1973},
  doi = {10.1080/00018737300101379}
}

@article{WaldWolfowitz1940,
  author = {Wald, A. and Wolfowitz, J.},
  title = {On a Test Whether Two Samples are from the Same Population},
  journal = {The Annals of Mathematical Statistics},
  volume = {11},
  number = {2},
  pages = {147--162},
  year = {1940},
  doi = {10.1214/aoms/1177731909}
}

@article{PelissettoVicari2002,
  author = {Pelissetto, Andrea and Vicari, Ettore},
  title = {Critical phenomena and renormalization-group theory},
  journal = {Physics Reports},
  volume = {368},
  number = {6},
  pages = {549--727},
  year = {2002},
  doi = {10.1016/S0370-1573(02)00219-3}
}

@article{BarkerHopfield1964,
  author = {Barker, A. S. and Hopfield, J. J.},
  title = {Coupled-Optical-Phonon-Mode Theory of the Infrared Dispersion in {BaTiO$_3$}, {SrTiO$_3$}, and {KTaO$_3$}},
  journal = {Physical Review},
  volume = {135},
  number = {6A},
  pages = {A1732--A1737},
  year = {1964},
  doi = {10.1103/PhysRev.135.A1732}
}
\end{ppsupplement}
\end{refsection}

\end{document}